\documentclass[12pt]{article}
\usepackage{graphicx}
\def\bra#1{\left\langle #1\right|}
\def\ket#1{\left| #1\right\rangle}
\newcommand{\bers}{\begin{eqnarray*}}
\newcommand{\eers}{\end{eqnarray*}}
\newcommand{\bt}{\begin{itemize}}
\newcommand{\et}{\end{itemize}}
\def\beq{\begin{equation}}
\def\eeq{\end{equation}}
\def\bea{\begin{eqnarray}}
\def\eea{\end{eqnarray}}
\def\nn{\nonumber}
\def\sla#1{\raise.15ex\hbox{$/$}\kern-.57em #1}% Feynman slash

\def\sss{\scriptscriptstyle}
\def\Bbar{{\bar B}^0}
\def\phiM{\phi_{\sss M}}
\def\abar{{\bar a}}
\def\bbar{{\bar b}}
\def\cbar{{\bar c}}
\def\bd{B_d^0}
\def\bdbar{{\overline{B_d^0}}}
\def\bs{B_s^0}
\def\bsbar{{\overline{B_s^0}}}
\def\barp{{\raise.35ex\hbox
{${\sss (}$}}---{\raise.35ex\hbox{${\sss )}$}}}
\def\bdbarp{\hbox{$B_d$\kern-1.4em\raise1.4ex\hbox{\barp}}}
\def\bsbarp{\hbox{$B_s$\kern-1.4em\raise1.4ex\hbox{\barp}}}
\def\ks{K_{\sss S}}

\def\roughly#1{\mathrel{\raise.3ex\hbox
{$#1$\kern-.75em\lower1ex\hbox{$\sim$}}}}

\def\epjc#1#2#3{{ Eur.\ Phys.\ J.}\ {\bf C#1}, #3, (#2)} 
\def\ijmp#1#2#3{{ Int.\ J.\ Mod.\ Phys.} {\bf A#1}, #3 (#2)}
\def\mpla#1#2#3{{Mod.\ Phys.\ Lett.} {\bf A#1}, #3 (#2)}
\def\nci#1#2#3{{ Nuovo Cimento} {\bf #1}, #3 (#2)}
\def\npb#1#2#3{{ Nucl.\ Phys.} {\bf B#1}, #3 (#2)}
\def\plb#1#2#3{{ Phys.\ Lett.} {\bf #1B}, #3 (#2)}
\def\prd#1#2#3{{ Phys.\ Rev.} {\bf D#1}, #3 (#2)}
\def\newprd#1#2#3{{ Phys.\ Rev.} {\bf D#1}, #3 (#2)}
\def\prl#1#2#3{{ Phys.\ Rev.\ Lett.} {\bf #1}, #3 (#2)}
\def\newprl#1#2#3{{ Phys.\ Rev.\ Lett.} {\bf #1}, #3 (#2)}
\def\zpc#1#2#3{{ Zeit.\ Phys.} {\bf C#1}, #3 (#2)}

\begin{document}

\begin{flushright}  
UdeM-GPP-TH-03-108\\
\end{flushright}

\begin{center}
\bigskip
{\Large \bf 
Triple-Product Correlations in $B\to V_1 V_2$ Decays and New
Physics
} \\
\bigskip
{\large Alakabha Datta $^{a,b,}$\footnote{datta@physics.utoronto.ca}
and David London $^{b,}$\footnote{london@lps.umontreal.ca}} \\
\end{center}

\begin{flushleft}
~~~~~~~~~~~$a$: {\it Department of Physics, University of Toronto,}\\
~~~~~~~~~~~~~~~{\it 60 St.\ George Street, Toronto, ON, Canada M5S 1A7}\\
~~~~~~~~~~~$b$: {\it Laboratoire Ren\'e J.-A. L\'evesque, 
Universit\'e de Montr\'eal,}\\
~~~~~~~~~~~~~~~{\it C.P. 6128, succ. centre-ville, Montr\'eal, QC,
Canada H3C 3J7}
\end{flushleft}
\begin{center} 
%\bigskip (\today)
\vskip0.5cm
{\Large Abstract\\}
\vskip3truemm

\parbox[t]{\textwidth} 
{In this paper we examine T-violating triple-product
correlations (TP's) in $B \to V_1 V_2$ decays. TP's are excellent
probes of physics beyond the standard model (SM) for two reasons: (i)
within the SM, most TP's are expected to be tiny, and (ii) unlike
direct CP asymmetries, TP's are not suppressed by the small strong
phases which are expected in $B$ decays. TP's are obtained via the
angular analysis of $B \to V_1 V_2$. In a general analysis based on
factorization, we demonstrate that the most promising decays for
measuring TP's in the SM involve excited final-state vector mesons,
and we provide estimates of such TP's. We find that there are only a
handful of decays in which large TP's are possible, and the size of
these TP's depends strongly on the size of nonfactorizable effects. We
show that TP's which vanish in the SM can be very large in models with
new physics. The measurement of a nonzero TP asymmetry in a decay
where none is expected would specifically point to new physics
involving large couplings to the right-handed $b$-quark.}
\end{center}

\thispagestyle{empty}
\newpage
\setcounter{page}{1}
% Decrease texheight (for preprint numbers) again
%\textheight 23.0 true cm
\baselineskip=14pt

\section{Introduction}

There is a great deal of interest these days in the study of CP
violation in the $B$ system. By examining CP-violating effects in $B$
decays, we hope to get some clues as to the origin of CP violation in
the quark sector. If we are lucky, the standard model (SM) explanation
of CP violation --- a complex phase in the Cabibbo-Kobayashi-Maskawa
(CKM) matrix --- will be shown to be insufficient to explain the data,
and we will therefore have found indirect evidence for the presence of
physics beyond the SM.

Most of the theoretical work on this subject has concentrated on
mixing-induced CP-violating asymmetries in neutral $B$ decays, while a
smaller fraction has focussed on direct CP asymmetries
\cite{CPreview}. However, there is another class of CP-violating
effects which has received considerably less attention, and which can
also reveal the presence of new physics: triple-product
correlations. These take the form $\vec v_1 \cdot (\vec v_2 \times
\vec v_3)$, where each $v_i$ is a spin or momentum. These triple
products (TP's) are odd under time reversal (T) and hence, by the CPT
theorem, also constitute potential signals of CP violation. One can
establish the presence of a nonzero TP by measuring a nonzero value of
the asymmetry
\beq
A_{\sss T} \equiv 
{{\Gamma (\vec v_1 \cdot (\vec v_2 \times \vec v_3)>0) - 
\Gamma (\vec v_1 \cdot (\vec v_2 \times \vec v_3)<0)} \over 
{\Gamma (\vec v_1 \cdot (\vec v_2 \times \vec v_3)>0) + 
\Gamma (\vec v_1 \cdot (\vec v_2 \times \vec v_3)<0)}} ~,
\label{Toddasym}
\eeq
where $\Gamma$ is the decay rate for the process in question.

Of course, there is a well-known technical complication for such
effects: strong phases can produce a nonzero value of $A_{\sss T}$,
even if the weak phases are zero (i.e.\ there is no CP violation).
Thus the TP asymmetry $A_{\sss T}$ is not a true T-violating effect
(we refer to it as {\it T-odd}). However, one can still obtain a true
T-violating (and hence CP-violating) signal by comparing $A_{\sss T}$
with ${\bar A}_{\sss T}$, where ${\bar A}_{\sss T}$ is the T-odd
asymmetry measured in the CP-conjugate decay process \cite{Valencia}.

TP asymmetries are similar to direct CP asymmetries in two ways: (i)
they are both obtained by comparing a signal in a given decay with the
corresponding signal in the CP-transformed process, and (ii) both are
nonzero only if there are two interfering decay amplitudes. However,
there is an important difference between the two. Denoting $\phi$ and
$\delta$ as the relative weak and strong phases, respectively, between
the two interfering amplitudes, the signal for direct CP violation can
be written
\beq
{\cal A}_{\sss CP}^{dir} \propto \sin\phi \sin\delta ~,
\label{Adirform}
\eeq
while, as we shall see, that for the (true T-violating) TP asymmetry
is given by
\beq
{\cal A}_{\sss T}  \propto \sin\phi \cos\delta ~.
\label{TPform}
\eeq
The key point here is that one can produce a direct CP asymmetry only
if there is a nonzero strong-phase difference between the two decay
amplitudes. However, it has been argued that, due to the fact that the
$b$-quark is heavy, all strong phases in $B$ decays should be rather
small. In this case, all direct CP-violation signals will be tiny as
well. On the other hand, TP asymmetries are {\it maximal} when the
strong-phase difference vanishes. Thus, it may well be more promising
to search for triple-product asymmetries than direct CP asymmetries in
$B$ decays.

One class of $B$ processes in which triple products are generally
expected to appear are the decays of a $B$-meson (charged or neutral)
into two final-state vector mesons: $B\to V_1 V_2$. In the rest frame
of the $B$, the TP takes the form ${\vec q} \cdot ({\vec\varepsilon}_1
\times {\vec\varepsilon}_2)$, where ${\vec q}$ is the momentum of one
of the final vector mesons, and ${\vec\varepsilon}_1$ and
${\vec\varepsilon}_2$ are the polarizations of $V_1$ and $V_2$. Since
$\bd$ and $B^\pm$ mesons are already being produced copiously at the
$B$-factories BaBar and Belle, the study of such TP signals can be
performed now.

Some triple-product signals in the $B$ system have been studied within
the SM in past analyses -- they were first examined many years ago by
Valencia \cite{Valencia}, and several general studies of $B\to V_1
V_2$ decays were subsequently performed \cite{KP,DDLR,CW,Chiang}. In
these papers, it is found that the TP's with ground state vector
mesons are (almost) all small. As we show in the present paper, this
result can be understood, in a general analysis based on
factorization, in terms of mass and flavour suppressions. On the other
hand, these suppressions are small or absent for decays involving
excited vector mesons. We therefore note that the most promising
decays for measuring TP's in the SM involve radially-excited mesons,
and we provide estimates of the TP's in such decays, as well as in
several other modes not considered previously. However, as we show,
most of these TP asymmetries are also expected to be small in the
SM. The fact that most TP's are expected to be small in the SM makes
their measurements a very promising method for searching for new
physics.

We begin in Sec.~2 with a general review of triple-product
correlations in $B\to V_1 V_2$ decays. Using factorization, we
describe the conditions which must be present in order to produce a TP
in a given decay. We then present a detailed list of exclusive decays
which are expected to yield TP's in the SM. We also discuss the
possibility of generating TP's via mixing. In Sec.~3, we turn to the
question of the experimental prospects for measuring TP's. It is well
known that one can disentangle the helicities of the $V_1 V_2$ final
state via an angular analysis. We briefly review this analysis,
stressing that this is precisely how TP's are measured. As we will
show, TP's are typically suppressed by a factor of at least $m_{\sss
V}/m_{\sss B}$, and are further suppressed if $V_1$ and $V_2$ are
related by a symmetry. Consequently, the TP's in $B$ decays to
ground-state vector mesons are all expected to be very small, and this
has been found by previous analyses. On the other hand, decays in
which the final-state vector mesons are unrelated, and as heavy as
possible, are less affected by these suppressions.  The most promising
decays for the detection of TP's are therefore those which involve
final-state radially-excited mesons. In this section, we estimate the
size of the TP's, as well as the branching ratios, for such decays.
Although there are some TP's which may be large, the great majority of
$B$ decays exhibit very small TP's. This makes them an ideal place to
look for physics beyond the SM -- should any large TP be found, this
would be a clear signal of the presence of new physics. We also
address the issue of nonfactorizable effects in this section, as well
as how TP's may help in the resolution of discrete ambiguities in the
measurement of the angles of the unitarity triangle. Finally, in
Sec.~4, we examine the properties of new physics which can modify the
SM predictions for TP's in various $B\to V_1 V_2$ decays.
Specifically, we show that if the new physics involves significant
couplings to the right-handed $b$-quark, large TP asymmetries can be
produced. We illustrate this in the context of a specific new-physics
model, supersymmetry with broken R-parity. This demonstrates quite
clearly that the measurement of TP's is an excellent way to search for
new physics. We summarize our results in Sec.~5.

\section{Triple Products in $B\to VV$ decays}

\subsection{General Considerations}

In this subsection, we follow the analysis of Ref.~\cite{Valencia},
and use the following notation at the meson level: $B(p) \to
V_1(k_1,\varepsilon_1) + V_2(k_2,\varepsilon_2)$. The decay amplitude
can then be expressed as follows:
\beq
M = a \, \varepsilon_1^* \cdot \varepsilon_2^* + {b \over m_B^2}
(p\cdot \varepsilon_1^*) (p\cdot \varepsilon_2^*) + i {c \over m_B^2}
\epsilon_{\mu\nu\rho\sigma} p^\mu q^\nu \varepsilon_1^{*\rho}
\varepsilon_2^{*\sigma} ~,
\label{abcdefs}
\eeq
where $q\equiv k_1 - k_2$. (Note that we have normalized terms with a
factor $m_B^2$, rather than $m_1 m_2$ as in Ref.~\cite{Valencia}.
With the above normalization, each of $a$, $b$ and $c$ is expected to
be the same order of magnitude.) The $a$ and $b$ terms correspond to
combinations of $s$- and $d$-wave amplitudes while the $c$ term
corresponds to the $p$-wave amplitude for the final state. The
quantities $a$, $b$ and $c$ are complex and will in general contain
both CP-conserving strong phases and CP-violating weak phases.

In $|M|^2$, a triple-product correlation arises from interference
terms involving the $c$ amplitude, and will be present if ${\rm Im}(a
c^*)$ or ${\rm Im}(b c^*)$ is nonzero. In the rest frame of the $B$
meson, this TP takes the form ${\vec q} \cdot ({\vec\varepsilon}_1^*
\times {\vec\varepsilon}_2^*)$.

However, as discussed above, due to the presence of strong phases,
such TP's are not necessarily true T-violating effects. To obtain a
true measure of T violation, one has to compare the triple product
measured in $B\to V_1 V_2$ with that obtained in the CP-conjugate
process. Using CPT, the amplitude for the CP-conjugate process
$\bar{B}(p) \to \bar{V}_1(k_1,\varepsilon_1) +
\bar{V}_2(k_2,\varepsilon_2)$ can be expressed as follows:
\beq
{\overline{M}} = \abar \, \varepsilon_1^* \cdot \varepsilon_2^* +
{\bbar \over m_B^2} (p\cdot \varepsilon_1^*) (p\cdot \varepsilon_2^*)
- i {\cbar \over m_B^2} \epsilon_{\mu\nu\rho\sigma} p^\mu q^\nu
\varepsilon_1^{*\rho} \varepsilon_2^{*\sigma} ~,
\label{abcdefsbar}
\eeq
where $\abar$, $\bbar$ and $\cbar$ can be obtained from $a$, $b$
and $c$ by changing the sign of the weak phases. If CP is conserved,
one has $\abar=a$, $\bbar=b$ and $\cbar=c$.

Note that CPT leaves invariant each of the three Lorentz scalars in
Eq.~(\ref{abcdefs}). Thus, because the $p$-wave amplitude in
${\overline{M}}$ changes sign relative to that of $M$, the sign of the
T-odd asymmetry in $|{\overline{M}}|^2$ is {\it opposite} that in
$|M|^2$. The true T-violating asymmetry is therefore found by {\it
adding} the T-odd asymmetries in $|M|^2$ and $|{\overline{M}}|^2$
\cite{Valencia}:
\beq
{\cal A}_{\sss T} \equiv {1\over 2}(A_T + {\bar A}_T) ~.
\label{Tviolasym}
\eeq
Writing
\bea
a = \sum_i a_i e^{i \phi_i^a} e^{i \delta_i^a} & ~~,~~~~ &
\abar = \sum_i a_i e^{-i \phi_i^a} e^{i \delta_i^a} ~, \\
b = \sum_i b_i e^{i \phi_i^b} e^{i \delta_i^b} & ~~,~~~~ & 
{\bar b} = \sum_i b_i e^{-i \phi_i^b} e^{i \delta_i^b} ~, \\
c = \sum_i c_i e^{i \phi_i^c} e^{i \delta_i^c} & ~~,~~~~ & 
{\bar c} = \sum_i c_i e^{-i \phi_i^c} e^{i \delta_i^c} ~,
\label{abcexplicit}
\eea
where the $\phi_i^{a,b,c}$ ($\delta_i^{a,b,c}$) are weak (strong)
phases, we see that
\bea
{1\over 2}\left[ {\rm Im}(a c^*) - {\rm Im}(\abar {\bar c}^*)
\right] & = & \sum_{i,j} a_i c_j \sin \left( \phi_i^a - \phi_j^c
\right) \cos \left( \delta_i^a - \delta_j^c \right) ~, \\
{1\over 2}\left[ {\rm Im}(b c^*) - {\rm Im}({\bar b} {\bar c}^*)
\right] & = & \sum_{i,j} b_i c_j \sin \left( \phi_i^b - \phi_j^c
\right) \cos \left( \delta_i^b - \delta_j^c \right) ~,
\label{TPform2}
\eea
which explains the form of Eq.~(\ref{TPform}).

\subsection{Factorization}

Not all $B\to V_1 V_2$ decays will necessarily yield triple products.
In this subsection, we use the framework of naive factorization to
examine the conditions which are required in order to produce a TP in
a given decay. It should be noted that there have been recent
developments in the study of nonleptonic decays, such as QCD
factorization \cite{BBNS} and PQCD \cite{PQCD}, in which corrections
to naive factorization proportional to $\alpha_s$ have been calculated
in the heavy $m_B$ limit. QCD factorization has been applied to some
$B \to V_1 V_2$ decays \cite{ChengYang}. However, some of the
corrections to naive factorization turn out to be divergent, so that
predictive power is lost.

Previous analyses, using naive factorization, have found that most TP
asymmetries with ground state vector mesons are expected to be small
in the SM \cite{Valencia,KP,DDLR,CW,Chiang}. As will be shown, we
agree with this result. Note that this conclusion will necessarily
hold even if one employs QCD factorization or PQCD, since the dominant
contribution comes from naive factorization in these approaches. It
is possible that nonfactorizable effects are significant in certain $B
\to V_1 V_2$ decays, particularly those dominated by colour-suppressed
amplitudes. We discuss these effects in some detail later (Sec.~3.3),
and attempt to take them into account in our analysis. For $B$ decays
to radially-excited vector mesons, which have not been considered
previously, we also use naive factorization to estimate the TP
asymmetries. (Note that the methods of QCD factorization or PQCD have
not been developed or used with radially-excited states.)

The starting point for factorization is the SM effective hamiltonian
for $B$ decays \cite{BuraseffH}:
\bea
H_{eff}^q &=& {G_F \over \protect \sqrt{2}}
[V_{fb}V^*_{fq}(c_1O_{1f}^q + c_2 O_{2f}^q) \nn\\
&& \qquad - \sum_{i=3}^{10}(V_{ub}V^*_{uq} c_i^u
+V_{cb}V^*_{cq} c_i^c +V_{tb}V^*_{tq} c_i^t) O_i^q] + h.c.,
\label{Heff}
\eea
where the superscript $u$, $c$, $t$ indicates the internal quark, $f$
can be the $u$ or $c$ quark, and $q$ can be either a $d$ or $s$ quark.
The operators $O_i^q$ are defined as
\bea
O_{f1}^q &=& \bar q_\alpha \gamma_\mu Lf_\beta\bar
f_\beta\gamma^\mu Lb_\alpha\;,\;\;\;\;\;\;O_{2f}^q =\bar q
\gamma_\mu L f\bar
f\gamma^\mu L b\;,\nn\\
O_{3,5}^q &=&\bar q \gamma_\mu L b
\bar q' \gamma^\mu L(R) q'\;,\;\;\;\;\;\;\;O_{4,6}^q = \bar q_\alpha
\gamma_\mu Lb_\beta
\bar q'_\beta \gamma^\mu L(R) q'_\alpha\;,\\
O_{7,9}^q &=& {3\over 2}\bar q \gamma_\mu L b  e_{q'}\bar q'
\gamma^\mu R(L)q'\;,\;O_{8,10}^q = {3\over 2}\bar q_\alpha
\gamma_\mu L b_\beta
e_{q'}\bar q'_\beta \gamma^\mu R(L) q'_\alpha ~, \nn
\eea
where $R(L) = 1 \pm \gamma_5$, and $q'$ is summed over $u$, $d$, $s$,
$c$. $O_2$ and $O_1$ are the tree-level and QCD-corrected operators,
respectively. $O_{3-6}$ are the strong gluon-induced penguin
operators, and operators $O_{7-10}$ are due to $\gamma$ and $Z$
exchange (electroweak penguins), and ``box'' diagrams at loop level.
In what follows, the important point is that all SM operators involve
a left-handed $b$-quark.

Within factorization, the amplitude for $B\to V_1 V_2$ can be written
as
\beq
{\cal A}(B \to V_1 V_2) = \sum_{{\cal O},{\cal O}'} \left\{ \bra{V_1}
{\cal O} \ket{0} \bra{V_2} {\cal O}' \ket{B} + \bra{V_2} {\cal O}
\ket{0} \bra{V_1} {\cal O}' \ket{B} \right\} ~,
\label{2amps}
\eeq
where ${\cal O}$ and ${\cal O}'$ are pieces of the $O^q_i$ operators
above. The specific quark content of these operators depends on the
final state $V_1 V_2$. (As mentioned above, we return to the question
of nonfactorizable effects in Sec.~3.)

In the following, we write the quark-level decay as $b\to q {\bar q}'
q'$, and call the spectator quark ${\bar q}_s$. As noted above, there
are two categories of operators which contribute to this decay: (i)
tree contributions, which have the form ${\bar q}' \gamma^\mu (1 -
\gamma_5) b \, {\bar q} \gamma_\mu (1 - \gamma_5) q'$, and (ii)
penguin operators, which have the form ${\bar q} \gamma^\mu (1 -
\gamma_5) b \, {\bar q}' \gamma_\mu (1 \pm \gamma_5) q'$. (In all
operators, both colour assignments are understood.)

Consider now the first term in the above expression, $\sum_{{\cal
O},{\cal O}'} \bra{V_1} {\cal O} \ket{0} \bra{V_2} {\cal O}' \ket{B}$.
Let us first suppose that $V_1 = q {\bar q}'$ (so-called
colour-allowed decays). Then
\beq
\bra{V_1} {\bar q} \gamma^\mu q' \ket{0} = m_1 g_{V_1}
\varepsilon_1^{*\mu} ~.
\eeq
The tree operator involves the factor ${\bar q} \gamma_\mu (1 -
\gamma_5) q'$, and therefore has the right form. On the other hand,
one must perform a Fierz transformation on the penguin operators to
obtain the correct form. Those operators of the form ${\bar q}
\gamma^\mu (1 - \gamma_5) b {\bar q}' \gamma_\mu (1 - \gamma_5) q'$
Fierz-transform into ${\bar q}' \gamma^\mu (1 - \gamma_5) b {\bar q}
\gamma_\mu (1 - \gamma_5) q'$, just like the tree contributions
(modulo colour factors). However, penguin operators of the form ${\bar
q} \gamma^\mu (1 - \gamma_5) b {\bar q}' \gamma_\mu (1 + \gamma_5) q'$
Fierz-transform into $-2{\bar q}' (1 - \gamma_5) b {\bar q} (1 +
\gamma_5) q'$, and these will not contribute to the decay, since
$\bra{V_1} {\bar q} (1 + \gamma_5) q' \ket{0} = 0$. We therefore find
that
\beq
\sum_{{\cal O},{\cal O}'} \bra{V_1} {\cal O} \ket{0} \bra{V_2} {\cal
O}' \ket{B} = X \varepsilon_1^{*\mu} \bra{V_2} {\bar q}' \gamma_\mu
(1 - \gamma_5) b \ket{B} ~,
\label{firstterm}
\eeq
where $X$ is a factor which includes a combination of Wilson
coefficients and weak CKM phases (e.g.\ $V_{cb} V_{cq}^*$ and $V_{tb}
V_{tq}^*$). The upshot is that, within the SM, there is only one decay
amplitude (i.e.\ operator) for this term, and the fundamental reason
for this is that the SM involves only left-handed $b$-quarks.

One obtains a similar expression for the case where $V_1 = q' {\bar
q}'$ (so-called colour-suppressed (or electroweak penguin) decays). The
only difference is that, in this case, the penguin operators have the
correct form, but the tree operator must be Fierz-transformed.
However, one still ends up with an expression for the amplitude
similar to that above.

Now, in order to cast Eq.~(\ref{firstterm}) in the same form as
Eq.~(\ref{abcdefs}), one must express the remaining matrix element
above in terms of form factors. This can be done as follows
\cite{BSW}:
\bea
\bra{V_2(k_2)} {\bar q}' \gamma_\mu b \ket{B(p)}& =& i {{2
V^{(2)}(r^2)} \over (m_B+m_{2})} \epsilon_{\mu \nu\rho\sigma} p^\nu
k_2^\rho \varepsilon_2^{*\sigma} ~,\nn\\
\bra{V_2(k_2)} {\bar q}' \gamma_\mu\gamma_5 b \ket{B(p)}& =&
(m_B+m_{2})A_1^{(2)}(r^2) \left[\varepsilon_{2\mu}^{*}-
\frac{\varepsilon_2^{*}.r}{r^2}r_{\mu}\right]\nn\\
&& ~-A_2^{(2)}(r^2)\frac{\varepsilon_2^{*}.r}{m_B+m_{2}}
\left[(p_{\mu}+k_{2\mu})- \frac{m_B^2-m_{2}^2}{r^2}r_{\mu}\right]\nn\\
&& ~+2 im_2\frac{\varepsilon_2^{*}.r}{r^2}r_{\mu}A_0^{(2)}(r^2) ~,
\label{ffactor}
\eea   
where $r=p-k_2$, and $V^{(2)}$, $A_1^{(2)}$, $A_2^{(2)}$ and
$A_0^{(2)}$ are form factors. Thus, the first term of
Eq.~(\ref{2amps}) is given by
\bea
\sum_{{\cal O},{\cal O}'} \bra{V_1} {\cal O} \ket{0} \bra{V_2} {\cal
O}' \ket{B}& =& -(m_B+m_2)m_1g_{V_1}XA_1^{(2)}(m_1^2)
 \varepsilon_1^* \cdot \varepsilon_2^* \nn\\ 
&& ~ +2\frac{m_1}{m_B+m_2}g_{V_1}X A_2^{(2)}(m_1^2) \varepsilon_2^*
\cdot p \varepsilon_1^* \cdot p \nn\\
&& ~ -i \frac{m_1}{ (m_B+m_2)}g_{V_1}X V^{(2)}(m_1^2)\epsilon_{\mu
\nu\rho\sigma} p^\mu q^\nu\varepsilon_1^{*\rho}
\varepsilon_2^{*\sigma} ~,
\label{term1}
\eea
where we have used $k_2=(p-q)/2$. The key point here is that all phase
information is contained within the factor $X$, which is common to all
three independent amplitudes. {\it Thus, these quantities all have the
same phase.}

A similar analysis holds for the second term in Eq.~(\ref{2amps}):
\bea
\sum_{{\cal O},{\cal O}'} \bra{V_2} {\cal O} \ket{0} \bra{V_1} {\cal
O}' \ket{B}& =& -(m_B+m_1)m_2g_{V_2}YA_1^{(1)}(m_2^2) \varepsilon_1^*
\cdot \varepsilon_2^* \nn\\
&& ~ +2\frac{m_2}{m_B+m_1}g_{V_2}Y A_2^{(1)}(m_2^2) \varepsilon_2^*
\cdot p \varepsilon_1^* \cdot p \nn\\
&& ~ -i \frac{m_2}{ (m_B+m_1)}g_{V_2}Y V^{(1)}(m_2^2)\epsilon_{\mu
\nu\rho\sigma} p^\mu q^\nu\varepsilon_1^{*\rho}
\varepsilon_2^{*\sigma} ~.
\label{term2} 
\eea
As before, all three independent amplitudes have the same phase, $Y$
(though this is not necessarily equal to that of the first term, $X$).

We can now express the quantities $a$, $b$ and $c$ of
Eq.~(\ref{abcdefs}) as follows:
\bea
a &=& -m_1g_{V_1}(m_B+m_2) A_1^{(2)}(m_1^2) X - m_2g_{V_2}(m_B+m_1)
A_1^{(1)}(m_2^2) Y\nn\\
b &=& 2m_1g_{V_1}{m_B \over (m_B+m_2)}m_B A_2^{(2)}(m_1^2) X +
2m_2g_{V_2}{m_B \over (m_B+m_1)}m_B A_2^{(1)}(m_2^2) Y\nn\\
c &=& -m_1g_{V_1}{m_B \over (m_B+m_2)}m_B V^{(2)}(m_1^2) X
-m_2g_{V_2}{m_B \over (m_B+m_1)}m_B V^{(1)}(m_2^2) Y ~.
\label{abc}
\eea

At this point we can make an important general observation. As noted
previously, TP's will be produced in $B \to V_1 V_2$ decays as long as
${\rm Im}(a c^*)$ or ${\rm Im}(b c^*)$ is nonzero. However, from the
above equation, we see that if either $X$ or $Y$ is zero, then $a$,
$b$ and $c$ will all have the same phase, so that ${\rm Im}(ac^*) =
{\rm Im}(bc^*) = 0$. Therefore, in order to have a triple-product
correlation in a given decay, {\it both of the amplitudes in
Eq.~(\ref{2amps}) must be present.}

This is perhaps a surprising result. Naively, one would think that if
a particular decay receives both tree and penguin contributions, with
different weak phases, T-violating TP's would automatically arise.
However, as we have shown above, this is not necessarily so. The
reason is that TP's are a {\it kinematical} CP-violating effect
\cite{Kayser}. It is therefore not enough to have two decay amplitudes
with a relative weak phase. What one really needs is two different
{\it kinematical} amplitudes with a relative weak phase.

There is a second important point: if we replace the index `2' by `1'
in Eq.~(\ref{abc}) above, $a$, $b$ and $c$ will once again have the
same phase. We therefore see explicitly that if $V_1 = V_2$, no TP's
can be produced. (This is to be expected since, from
Eq.~(\ref{2amps}), there is only a single amplitude in this case.)
However, it also indicates that if $V_1$ and $V_2$ are similar, i.e.\
related by a symmetry, the phases of $a$, $b$ and $c$ will also be
similar, and the TP correspondingly suppressed. This will be important
when we estimate the sizes of TP's for specific exclusive decays.

\subsection{Triple products in specific exclusive decays}

We now turn to establishing which specific exclusive $B \to V_1 V_2$
decays are expected to have triple-product correlations in the SM. As
we have noted above, two kinematical amplitudes are necessary in order
for a TP to be produced.

To a first approximation, there is a simple rule for determining which
processes have two such amplitudes: in the quark-level decay $b\to q
{\bar q}' q'$, if the spectator quark ${\bar q}_s$ is the same as
${\bar q}'$, then the two amplitudes of Eq.~(\ref{2amps}) will be
present. However, this is not sufficient to generate a true
T-violating TP. It is also necessary that the two kinematical
amplitudes have different weak phases. Thus, if the quark-level decay
is dominated by a single decay amplitude, a TP can never be
generated. This is the case for the quark-level decays $b \to c {\bar
c} s$, whose tree and penguin contributions have approximately the
same weak phase (an example of such a decay at the meson level is $B
\to J/\psi K^*$). It also holds for pure $b\to s$ penguin decays,
which are dominated by internal $t$-quark exchange (e.g.\ $B \to \phi
K^*$).

This rule must be modified slightly when the flavour wavefunction of
one of the final-state vector mesons contains more than one piece
(e.g.\ the $\rho^0$ is composed of both $u{\bar u}$ and $d{\bar d}$
pairs). In this case, several different quark-level $b\to q {\bar q}'
q'$ decays can contribute to the final state, and ${\bar q}_s$ must be
the same as one of the ${\bar q}'$ quarks. Furthermore, it is
necessary that $V_1$ and $V_2$ have different flavour wavefunctions.
For example, suppose that $V_1 = \rho^0$ and $V_2 = \rho^{0\prime}$,
where $\rho^{0\prime}$ is an excited state. In this case, even though
$V_1 \ne V_2$, there will still be no TP since the two kinematical
amplitudes will have the same phase, i.e.\ one will have $X=Y$ in
Eq.~(\ref{abc}).

With these constraints, we find that only a small number of $B$ decays
can yield TP's in the SM. These are listed below. In the discussion of
each decay mode, we use the following notation to denote the main
decay amplitudes: $T$ (colour-allowed tree amplitude), $C$
(colour-suppressed tree amplitude), $P$ (gluon-mediated penguin
amplitude), and $P_{EW}$ (electroweak penguin amplitude). We ignore
the smaller decay amplitudes such as the OZI-suppressed gluonic
penguin and the colour-suppressed electroweak penguin amplitudes
(although they will be included in our numerical calculations in the
next section). We also note which CKM matrix elements govern each of
the decay amplitudes. For $T$, $C$, $P_{EW}$ and $b \to s$ $P$
amplitudes these CKM elements are always well-defined. On the other
hand, $b\to d$ penguin amplitudes receive contributions from internal
$u$, $c$ and $t$ quarks, which involve different combinations of CKM
matrix elements. Using the unitarity of the CKM matrix, this amplitude
can always be written in terms of a piece proportional to $V_{tb}
V_{td}^*$ and another piece proportional to either $V_{ub} V_{ud}^*$
or $V_{cb} V_{cd}^*$. However, for most decays of interest, this
second piece can always be absorbed into a $T$ or $C$ amplitude. Thus,
the $b \to d$ penguin amplitudes can usually be thought of as
effectively governed by $V_{tb} V_{td}^*$ (there is one exception,
noted below). Note also that the final-state mesons in the list below
can be in the ground state or in an excited state.
\begin{itemize}

\item $B_c^- \to J/\psi D^{*-}$ ($b \to c {\bar c} d$). There are four
  contributing decay amplitudes: $T$ and $C$ ($V_{cb} V_{cd}^*$), and
  $P$ and $P_{EW}$ ($V_{tb} V_{td}^*$). $T$ and $P$ are kinematically
  similar, as are $C$ and $P_{EW}$. Thus, TP's arise from the
  interference of $T$ and $P_{EW}$ or $C$ and $P$.

\item $B^- \to \rho^0 K^{*-}$, $\omega K^{*-}$ ($b \to u {\bar u}
  s$). There are 4 contributing decay amplitudes: $T$, $C$ ($V_{ub}
  V_{us}^*$) and $P$ and $P_{EW}$ ($V_{tb} V_{ts}^*$). TP's arise from
  the interference of $T$ and $P_{EW}$ or $C$ and $P$.

\item $\bdbar \to K^{*0} \rho^0$, $K^{*0} \omega$. This is more
  complicated. There are 3 contributing amplitudes: $P$ and $P_{EW}$
  ($V_{tb} V_{ts}^*$) arise from the quark-level decay $b \to d {\bar
  d} s$, while $C$ ($V_{ub} V_{us}^*$) comes from $b \to u {\bar u}
  s$. The TP comes from the interference of $C$ and $P$.

\item $B^- \to \rho^- \rho^0$, $\rho^- \omega$. This is the most
  complicated decay. Here there are 4 contributing amplitudes. $T$
  and $C$ ($V_{ub} V_{ud}^*$) correspond to the quark-level decay $b
  \to u {\bar u} d$, while $P$ and $P_{EW}$ ($V_{tb} V_{td}^*$) come
  from both $b \to u {\bar u} d$ and $b \to d {\bar d} d$. However,
  the penguin decays $b \to u {\bar u} d$ and $b \to d {\bar d} d$ are
  kinematically different. The TP arises mainly from the interference
  of $T$ and the $b \to d {\bar d} d$ $P$, but other interferences can
  also contribute.

\item $\bdbar \to \rho^0 \omega$. There can be a TP in this decay
  due to the fact that $\rho^0$ and $\omega$ have different flavour
  wavefunctions: $\rho^0 = (u{\bar u} - d{\bar d})/\sqrt{2}$,
  $\omega = (u{\bar u} + d{\bar d})/\sqrt{2}$. There are several
  contributions: $C$ ($b\to u{\bar u}d$: $V_{ub} V_{ud}^*$), $P$ ($b
  \to d {\bar d} d$: $V_{tb} V_{td}^*$) and $P_{EW}$ ($b \to u {\bar
  u} d$ and $b \to d {\bar d} d$: $V_{tb} V_{td}^*$). The TP is due
  mainly to $C$--$P$ interference.

\item $\bsbar \to \phi K^{*0}$ ($b \to s {\bar s} d$). This is a pure
  penguin decay, and there are 2 contributing amplitudes: $P$ and
  $P_{EW}$. In this case, the TP arises from the interference of the
  $V_{tb} V_{td}^*$ and $V_{ub} V_{ud}^*$ pieces of these two $b\to d$
  penguin amplitudes.

\end{itemize}
Some of these decays have been studied previously: except for the
$B_c^-$ decays, the other decays in the SM, with the final-state
vector mesons in the ground state, have been examined in
Ref.~\cite{KP}.

In addition, there is another class of $B\to V_1 V_2$ decays, not
considered in earlier calculations, which can potentially yield
triple-product correlations:
\begin{itemize}

\item $B^- \to D^{*0} K^{*-}$ ($b \to c {\bar u} s$) receives
   contributions from $T$ and $C$, while $B^- \to {\bar D}^{*0}
   K^{*-}$ ($b \to u {\bar c} s$) is due to $C$ alone.

\item $B^- \to D^{*0} \rho^-$ ($b \to c {\bar u} d$) receives
   contributions from $T$ and $C$, while $B^- \to {\bar D}^{*0}
   \rho^-$ ($b \to u {\bar c} d$) is due to $C$ alone.

\item $B_c^- \to {\bar D}^{*0} D_s^{*-}$ ($b \to u {\bar c} s$)
   receives contributions from $T$ and $C$, while $B_c^- \to D^{*0}
   D_s^{*-}$ ($b \to c {\bar u} s$) is due to $C$ alone.

\item $B_c^- \to {\bar D}^{*0} D^{*-}$ ($b \to u {\bar c} d$)
   receives contributions from $T$ and $C$, while $B_c^- \to D^{*0}
   D^{*-}$ ($b \to c {\bar u} d$) is due to $C$ alone.

\end{itemize}
The main decay modes of the $D^{*0}$ are $D^0 \pi^0$ (60\%) and $D^0
\gamma$ (40\%). Similarly, the ${\bar D}^{*0}$ decays to ${\bar D}^0
\pi^0$ and ${\bar D}^0 \gamma$. Therefore, if we consider a final
state to which both $D^0$ and ${\bar D}^0$ can decay, the two
amplitudes in each of the items above can contribute to this decay. In
this case, the interference of $T$ ($b \to c {\bar u} f$, $f=d,s$) and
$C$ ($b \to u {\bar c} f$) will lead to a TP. (These types of
interferences are similar to those proposed for the extraction of the
CP phase $\gamma$ \cite{BDK,ADS}.)

In Sec.~3, we will provide estimates of the expected size of the TP's
within the SM for decays which have not been studied previously,
namely charmless $\bdbar$, $B^-$ and $\bsbar$ decays to final-state
excited vector mesons, $B_c^-$ decays, and $B^-$ and $B_c^-$ decays to
final states which include $D^0$ and ${\bar D}^0$ mesons.  Estimates
of TP's for charmless $\bdbar$ and $B^-$ decays to ground-state mesons
have already been given in Ref.~\cite{KP} and will not be repeated
here. We will, however, provide general arguments of why TP's with
ground-state mesons in the SM are small.

\subsection{Mixing-induced triple products}

Finally, there is one more possibility which must be examined. A decay
such as $\bdbar \to D^{*+} D^{*-}$ is not expected to yield a
triple-product correlation because, while the transition $\bdbar \to
D^{*+}$ is allowed, $\bdbar \to D^{*-}$ is not. That is, not both of
the amplitudes in Eq.~(\ref{2amps}) are present. However, the missing
amplitude can be generated via $\bd$--$\bdbar$ mixing: $\bdbar \to \bd
\to D^{*-}$. Thus, one might wonder whether this can lead to a TP. In
addition, even if a TP is expected in $B^0 \to V_1 V_2$, if $\Bbar$
can also decay to $V_1 V_2$, the TP may be modified in time due to
$B^0$--$\Bbar$ mixing. In this subsection, we investigate these
possibilities, which were first examined by Valencia in
Ref.~\cite{Valencia}.

In the presence of $B^0$--$\Bbar$ mixing, the states $B^0$ and $\Bbar$
can be written as a function of time as follows:
\bea
B^0(t) & = & e^{-i \left( M - i{\Gamma \over 2} \right) t} \left[ \cos
  \left({\Delta m \, t \over 2} \right) B^0 - i \, e^{-2 i \phiM}
  \sin \left({\Delta m \, t \over 2} \right) \Bbar \right], \nn\\
\Bbar(t) & = & e^{-i \left( M - i{\Gamma \over 2} \right) t} \left[ -
  i \, e^{2 i \phiM} \sin \left({\Delta m \, t \over 2} \right)
  B^0 + \cos \left({\Delta m \, t \over 2} \right) \Bbar \right],
\label{Btimedepamps}
\eea
where $\phiM$ is the weak phase in $B^0$--$\Bbar$ mixing [$\phiM =
\beta$ (0) for $B^0 = \bd$ ($\bs$)]. Following Eq.~(\ref{abcdefs}), we
define \cite{Valencia}
\bea
A(B^0 \to V_1 V_2) & = & a_1 s + b_1 d + i c_1 p \nn\\
A(\Bbar \to V_1 V_2) & = & a_2 s + b_2 d + i c_2 p \nn\\
A(\Bbar \to {\bar V}_1 {\bar V}_2) & = & \abar_1 s + \bbar_1 d - i
  \cbar_1 p \nn\\
A(B^0 \to {\bar V}_1 {\bar V}_2) & = & \abar_2 s + \bbar_2 d - i
  \cbar_2 p ~,
\label{BVVamps}
\eea
where $s$, $d$ and $p$ are defined as in Eq.~(\ref{abcdefs}): $s
\equiv \varepsilon_1^* \cdot \varepsilon_2^*$, $d \equiv (p\cdot
\varepsilon_1^*) (p\cdot \varepsilon_2^*) / m_B^2$, $p \equiv
\epsilon_{\mu\nu\rho\sigma} p^\mu q^\nu \varepsilon_1^{*\rho}
\varepsilon_2^{*\sigma} / m_B^2$. In the above, the barred amplitudes
are obtained from the corresponding unbarred ones by changing the sign
of the weak phases. We can then write
\beq
M \equiv A(B^0(t) \to V_1 V_2) = e^{-i \left( M - i{\Gamma \over 2}
\right) t} \left[ a s + b d + i c p \right],
\eeq
with
\bea
a & = & a_1 \cos \left({\Delta m \, t \over 2} \right) - i \, e^{-2 i
  \phiM} \sin \left({\Delta m \, t \over 2} \right) a_2 ~, \nn\\
b & = & b_1 \cos \left({\Delta m \, t \over 2} \right) - i \, e^{-2 i
  \phiM} \sin \left({\Delta m \, t \over 2} \right) b_2 ~, \nn\\
c & = & c_1 \cos \left({\Delta m \, t \over 2} \right) - i \, e^{-2 i
  \phiM} \sin \left({\Delta m \, t \over 2} \right) c_2 ~.
\eea
Similarly,
\beq
{\bar M} \equiv A(\Bbar(t) \to {\bar V}_1 {\bar V}_2) = e^{-i
\left( M - i{\Gamma \over 2} \right) t} \left[ \abar s + \bbar d - i
\cbar p \right],
\eeq
with
\bea
\abar & = & \abar_1 \cos \left({\Delta m \, t \over 2} \right) - i \,
  e^{2 i \phiM} \sin \left({\Delta m \, t \over 2} \right) \abar_2 ~,
  \nn\\
\bbar & = & \bbar_1 \cos \left({\Delta m \, t \over 2} \right) - i \,
  e^{2 i \phiM} \sin \left({\Delta m \, t \over 2} \right) \bbar_2 ~,
  \nn\\
\cbar & = & \cbar_1 \cos \left({\Delta m \, t \over 2} \right) - i \,
  e^{2 i \phiM} \sin \left({\Delta m \, t \over 2} \right) \cbar_2 ~.
\eea

Now, since we are interested in TP's, we will consider only the
$a$--$c$ interference terms in $|M|^2$ and $|{\bar M}|^2$ (the
conclusions will be identical for $b$--$c$ interference)
\cite{Chiang}. The T-violating term which interests us is found in the
sum of $|M|^2$ and $|{\bar M}|^2$ \cite{Valencia}:
\bea
|M|^2_{ac} + |{\bar M}|^2_{ac} & \sim & {\rm Im}(a \, c^*) - {\rm
 Im}(\abar \, \cbar^*) \nn\\
& = & \cos^2 \left({\Delta m \, t \over 2} \right) {\rm Im}(a_1 \,
c_1^* - \abar_1 \, \cbar_1^*) + \sin^2 \left({\Delta m \, t \over 2}
\right) {\rm Im}(a_2 \, c_2^* - \abar_2 \, \cbar_2^*) \nn\\
& & ~ + \sin \left({\Delta m \, t \over 2} \right) \cos
\left({\Delta m \, t \over 2} \right) {\rm Re} \left[e^{-2 i \phiM}
a_2 \, c_1^* - e^{2 i \phiM} \abar_2 \, \cbar_1^* \right. \nn\\
& & \hskip2.2truein \left.
- e^{2 i \phiM} a_1 \, c_2^* + e^{-2 i \phiM} \abar_1 \, \cbar_2^*
  \right].
\label{timedepTP}
\eea
The first term above is nonzero only if there is a TP in $B^0 \to V_1
V_2$, and describes how this TP evolves in time (note that it is the
only term which does not vanish at $t = 0$). Similarly, the second
term, which is generated due to $B^0$--$\Bbar$ mixing, describes the
time evolution of the TP in $\Bbar \to V_1 V_2$. Note that if the
final state is self-conjugate, ${\bar V}_1 {\bar V}_2 = V_1 V_2$, we
have [see Eq.~(\ref{BVVamps})]
\beq
\abar_2 = a_1 ~~,~~~~ \abar_1 = a_2 ~~,~~~~ \bbar_2 = b_1 ~~,~~~~
\bbar_1 = b_2 ~~,~~~~ \cbar_2 = -c_1 ~~,~~~~ \cbar_1 = -c_2 ~.
\label{selfconj}
\eeq
In this case, the first two terms of Eq.~(\ref{timedepTP}) add, and
the third term vanishes, so that the TP in $B^0 \to V_1 V_2$ is
independent of time.

Now consider the third term in Eq.~(\ref{timedepTP}). This is the term
which can potentially generate a TP due to $B^0$--$\Bbar$ mixing even
if the TP in $B^0 \to V_1 V_2$ is absent. Perhaps the easiest way to
see what is happening here is to explicitly write the amplitudes
$a_1$, $a_2$, etc.\ as in Eq.~(\ref{abcexplicit}):
\bea
a_1 = \sum_i a_{1i} e^{i \phi_i^{a_1}} e^{i \delta_i^{a_1}} & ~~,~~~~ &
\abar_1 = \sum_i a_{1i} e^{-i \phi_i^{a_1}} e^{i \delta_i^{a_1}} ~,\nn\\
a_2 = \sum_i a_{2i} e^{i \phi_i^{a_2}} e^{i \delta_i^{a_2}} & ~~,~~~~ &
\abar_2 = \sum_i a_{2i} e^{-i \phi_i^{a_2}} e^{i \delta_i^{a_2}} ~,\nn\\
c_1 = \sum_i c_{1i} e^{i \phi_i^{c_1}} e^{i \delta_i^{c_1}} & ~~,~~~~ &
\cbar_1 = \sum_i c_{1i} e^{-i \phi_i^{c_1}} e^{i \delta_i^{c_1}} ~,\nn\\
c_2 = \sum_i c_{2i} e^{i \phi_i^{c_2}} e^{i \delta_i^{c_2}} & ~~,~~~~ &
\cbar_2 = \sum_i c_{2i} e^{-i \phi_i^{c_2}} e^{i \delta_i^{c_2}} ~.
\eea
Then the third term can be written as
\bea
& & (\sin \Delta m t) \sum_{i,j} \left[ a_{2 i} c_{1 j} \sin(
  \phi_i^{a_2} - \phi_j^{c_1} - 2 \phiM) \sin(\delta_i^{a_2} -
  \delta_j^{c_1}) \right. \nn\\
& & \hskip1truein \left.
  - a_{1 i} c_{2 j} \sin( \phi_i^{a_1} - \phi_j^{c_2} + 2 \phiM)
  \sin(\delta_i^{a_1} - \delta_j^{c_2}) \right].
\label{mixingTP}
\eea
There are several points to be discussed here. It is indeed possible
to generate a T-violating triple product via $B^0$--$\Bbar$ mixing
even if the TP in $B^0 \to V_1 V_2$ is absent (note that we disagree
with Ref.~\cite{Valencia} on this point). However, unlike TP's
generated directly [e.g.\ Eq.~(\ref{TPform2})], these mixing-induced
TP's are similar to direct CP asymmetries in that they vanish when the
strong-phase differences vanish. Mathematically, the reason for this
can be traced to the factor of $i$ in the expression for the
time-dependent $B^0$ and $\Bbar$ states [Eq.~(\ref{Btimedepamps})].
But this can also be understood physically. As mentioned above, if the
transition $B^0 \to V_1$ is allowed, but $B^0 \to V_2$ is not, there
will be no TP. {}From Eq.~(\ref{mixingTP}), it appears that one can
generate a TP through $B^0$--$\Bbar$ mixing if $\Bbar \to V_2$ is
allowed. However, as we have stressed several times, TP's are
kinematical CP-violating effects. That is, we do not expect to
generate any TP's when the kinematics of the two amplitudes are the
same. Thus, the TP will still vanish if $\Bbar \to V_2$ is
kinematically identical to $B^0 \to V_1$. Since the kinematics are
related in part to the strong phases, it is not surprising that
mixing-induced TP's vanish when the strong-phase differences vanish.

In fact, this point can be quantified. Suppose the final state $V_1
V_2$ is self-conjugate, in which case the amplitudes satisfy the
relations in Eq.~(\ref{selfconj}). It is then straightforward to show
that the TP asymmetry described by Eq.~(\ref{mixingTP}) vanishes!
Thus, for example, even when $\bd$--$\bdbar$ mixing is taken into
account, one can never generate a TP in the decay $\bdbar \to D^{*+}
D^{*-}$, mentioned at the beginning of this subsection, because the
final state is self-conjugate. (That is, the transition $\bd \to
D^{*-}$ is kinematically identical to $\bdbar \to D^{*+}$, so that
$\bd$--$\bdbar$ mixing cannot lead to a TP.)

In light of this, we can now elaborate the conditions for generating a
TP via $B^0$--$\Bbar$ mixing: (i) the final state $V_1 V_2$ must be
one to which both $B^0$ and $\Bbar$ can decay, and (ii) it must not be
self-conjugate. Decays for which mixing can generate a TP include
$\bd\to D^{*+} D^{-\prime}$, $\bs \to K^{*-} K^{+\prime}$, $\bd\to
D^{*+} \rho^-$ \cite{D*rho}, etc., where $D^{-\prime}$ and
$K^{+\prime}$ are excited states.

Still, as noted above, this class of TP's is very similar to direct CP
asymmetries in that both involve the quantity $\sin \phi \sin\delta$
[see Eq.~(\ref{Adirform})]. Thus, compared to direct CP asymmetries,
we do not get additional information from these TP's. For this reason
we will not consider them further.

\section{Experimental Prospects}

In the previous section, we found several $B$ decays which are
predicted to exhibit triple-product correlations in the SM. The
relevant question now is: what are the prospects for detecting such
TP's experimentally? There are several issues here. What are the
experimental signals for TP's? For a given decay, what is the
branching ratio, and what is the expected size of the TP? In this
section, we provide answers to these questions.

\subsection{Experimental Signals}

In order to obtain experimental information from $B\to V_1 V_2$, it is
necessary to perform an angular analysis. For this purpose, it is
useful to use the linear polarization basis. In this basis, one
decomposes the decay amplitude into components in which the
polarizations of the final-state vector mesons are either longitudinal
($A_0$), or transverse to their directions of motion and parallel
($A_\|$) or perpendicular ($A_\perp$) to one another. One writes
\cite{DDLR,CW}
\beq
M = A_0 \varepsilon_1^{*\sss L} \cdot \varepsilon_2^{*\sss L} 
- {1 \over \sqrt{2}} A_\| {\vec\varepsilon}_1^{*\sss T} \cdot
  {\vec\varepsilon}_2^{*\sss T}
- {i \over \sqrt{2}} A_\perp {\vec\varepsilon}_1^{*\sss T} \times
  {\vec\varepsilon}_2^{*\sss T} \cdot {\hat p} ~,
\eeq
where ${\hat p}$ is the unit vector along the direction of motion of
$V_2$ in the rest frame of $V_1$, $\varepsilon_i^{*\sss L} =
{\vec\varepsilon}_i^* \cdot {\hat p}$, and ${\vec\varepsilon}_i^{*\sss
T} = {\vec\varepsilon}_i^* - \varepsilon_i^{*\sss L} {\hat p}$. $A_0$,
$A_\|$, $A_\perp$ are related to $a$, $b$ and $c$ of
Eq.~(\ref{abcdefs}) via
\beq
A_\| = \sqrt{2} a ~,~~~ A_0 = -a x - {m_1 m_2 \over m_{\sss B}^2} b
(x^2 - 1) ~,~~~ A_\perp = 2\sqrt{2} \, {m_1 m_2 \over m_{\sss B}^2} c 
\sqrt{x^2 - 1} ~,
\label{Aidefs}
\eeq
where $x = k_1 \cdot k_2 / (m_1 m_2)$. (A popular alternative basis is
to express the decay amplitude in terms of helicity amplitudes
$A_\lambda$, where $\lambda = 1, 0 ,-1$ \cite{KP,DDLR}. The helicity
amplitudes can be written in terms of the linear polarization
amplitudes via $A_{\pm 1} = (A_\| \pm A_\perp)/\sqrt{2}$, with $A_0$
the same in both bases.)

The angular distribution of the decay depends on the decay products of
$V_1$ and $V_2$. For the case where both vector mesons decay into
pseudoscalars, i.e.\ $V_1 \to P_1 P_1'$, $V_2 \to P_2 P_2'$, one has
\cite{DDLR,CW}
\bea
{d\Gamma \over d\cos\theta_1 d\cos\theta_2 d\phi} & = & N \left(
|A_0|^2 \cos^2\theta_1 \cos^2\theta_2 + {|A_\perp|^2 \over 2}
\sin^2\theta_1 \sin^2\theta_2 \sin^2 \phi \right. \nn\\
& & \hskip-1.0truein 
+ {|A_{\|}|^2 \over 2} \sin^2\theta_1 \sin^2\theta_2 \cos^2
\phi + {{\rm Re}(A_0 A_\|^*) \over 2\sqrt{2}} \sin 2\theta_1 \sin
2\theta_2 \cos\phi \nn\\
& &  \hskip-1.0truein \left. 
- {{\rm Im}(A_\perp A_0^*) \over 2\sqrt{2}} \sin 2\theta_1
\sin 2\theta_2 \sin\phi - {{\rm Im}(A_\perp A_\|^*) \over 2}
\sin^2\theta_1 \sin^2\theta_2 \sin 2\phi \right),
\label{angdist}
\eea
where $\theta_1$ ($\theta_2$) is the angle between the directions of
motion of the $P_1$ ($P_2$) in the $V_1$ ($V_2$) rest frame and the
$V_1$ ($V_2$) in the $B$ rest frame, and $\phi$ is the angle between
the normals to the planes defined by $P_1 P_1'$ and $P_2 P_2'$ in the
$B$ rest frame. (For other decays of the $V_1$ and $V_2$ (e.g.\ into
$e^+ e^-$, $P \gamma$ or three pseudoscalars), one will obtain a
different angular distribution, see Refs.~\cite{KP,DDLR,CW}.)

Now, the above angular distribution already appears in most of the
papers in Refs.~\cite{KP,DDLR,CW}. We repeat it here to emphasize the
following point. The terms which are of interest to us are those
proportional to ${\rm Im}(A_\perp A_0^*)$ and ${\rm Im}(A_\perp
A_\|^*)$. {}From Eq.~(\ref{Aidefs}) above, these are related to ${\rm
Im}(a c^*)$ and ${\rm Im}(b c^*)$. In other words, these two terms in
Eq.~(\ref{angdist}) are precisely the triple-product correlations.
Thus, by performing a full angular analysis, one can in fact obtain
the TP's.

Note that these terms are often referred to as CP-violating in
Refs.~\cite{KP,DDLR,CW}. However, as we have already noted, this is
not accurate -- they are really T-odd terms, and it is only by adding
the TP's in $|M|^2$ and $|{\overline{M}}|^2$ that one can obtain a
truly T-violating effect.

\subsection{Sizes of Triple Products -- Factorization}

In this subsection we estimate the sizes of the triple products within
factorization. We concentrate on those TP's which are generated
directly (i.e.\ not via mixing) because they do not vanish when the
strong phases vanish. Note: from the point of view of searching for
new physics, the precise predicted value of a given TP is not
particularly important. What is relevant is the question of whether
that TP is measurable ($>5\%$) or not. If it is expected to be small
within the SM, then the measurement of a large value for that TP would
point clearly towards the presence of physics beyond the SM. As we
will see, most TP's are expected to be very small in the SM.

As mentioned in the previous subsection, the presence of the terms
${\rm Im}(A_\perp A_0^*)$ or ${\rm Im}(A_\perp A_\|^*)$ in the angular
distribution will indicate a nonzero TP asymmetry. In order to
estimate the size of T violation in a given decay, we define the
following T-odd quantities:
\beq
A_T^{(1)} \equiv \frac{{\rm Im}(A_\perp
A_0^*)}{A_0^2+A_\|^2+A_\perp^2} ~~,~~~~
A_T^{(2)} \equiv \frac{{\rm Im}(A_\perp
A_\|^*)}{A_0^2+A_\|^2+A_\perp^2} ~.
\label{TPmeasure}
\eeq 
The corresponding quantities for the charge-conjugate process,
$\bar{A}_T^{(1)}$ and $\bar{A}_T^{(2)}$, are defined similarly. The
comparison of the TP asymmetries in a decay and in its corresponding
CP-conjugate process will give a measure of the true T-violating
asymmetry for that decay.

In order to calculate the TP quantities defined above, we first need
the values of $a$, $b$ and $c$ in Eq.~(\ref{abc}). These are obtained
using estimates of form factors, along with the latest Wilson
coefficients (including strong phases), decay constants and CKM matrix
elements. {}From these, we then calculate the linear polarization
amplitudes of Eq.~(\ref{Aidefs}) to obtain the branching ratios (BR's)
and T-odd TP's. As indicated above, in order to get true T-violating
TP's, we need to calculate $\abar$, $\bbar$ and $\cbar$ as well.

\subsection{SM Triple Products in $B$ Decays to Ground-State Vector Mesons}

As mentioned earlier, the sizes of TP's for $B$ decays to ground-state
vector mesons in the SM have been already estimated for many modes
\cite{KP}, and have been found to be small. Without performing any
actual calculations, we can understand this result by making some
general observations. First, as noted at the end of Sec.~2.3, if $V_1$
and $V_2$ are the same particle, the TP vanishes. Similarly, if $V_1 =
V_2$ in some symmetry limit, then there is again no TP in this limit
since $a$, $b$ and $c$ of Eq.~(\ref{abc}) are all proportional to $X +
Y$, and there is no relative phase. Thus, in this case the size of the
TP is related to the size of the symmetry breaking. We call this {\it
flavour suppression}. For example, we expect the TP in $B^- \to \rho^-
\rho^0$ to be tiny because the $\rho^-$ and $\rho^0$ are related by
isospin. Similarly, the TP in $B \to K^* \rho$ is expected to be small
since the $K^*$ and $\rho$ are related by flavour $SU(3)$ symmetry.

Second, consider $B \to V_1 V_2$ decays in which the final vector
mesons are light: $m_{1,2} \ll m_B$. Neglecting terms of
$O(m_{1,2}^2/m_B^2)$, we can then approximate $E_1 \sim E_2 \sim
|\vec{k}|= E = m_B/2$. Then, using Eq.~(\ref{Aidefs}), we have for the
various linear polarization amplitudes
\bea
A_0 & \approx & -(2a+b) \frac{E^2}{m_1m_2} ~, \nn\\
A_\| & \approx & \sqrt{2}a ~, \nn\\
A_\perp & \approx & \sqrt{2}c ~.
\label{smallmass}
\eea
Naively, since $a$, $b$ and $c$ are expected to be of the same order,
this then implies that
\beq
{A_{\|,\perp} \over A_0} \sim {m_1m_2 \over {E^2}} ~,
\label{suppression}
\eeq
We therefore expect the TP effects in $A_T^{(1)}$ to be suppressed by
$m_1 m_2 / {E^2}$, while those in $A_T^{(2)}$ are even smaller:
$A_T^{(2)} \sim (m_1 m_2 / {E^2})^2$. This behavior can be understood
rather simply. The form of the TP term in Eq.~(\ref{abcdefs}) requires
that both $V_1$ and $V_2$ be transversely polarized. However, the
polarization vector for transverse polarization is suppressed relative
to that for longitudinal polarization by $m/E$. This leads to the
above suppression factors for TP's in $B \to V_1 V_2$. (This to be
contrasted with T violation in $\Lambda_b \to F_1 V$ decays, where
$F_1$ is a spin-$1/2$ baryon and $V$ a vector meson. Here the
final-state $V$ can be longitudinally polarized, so that a T-violating
asymmetry can be produced without any suppression by powers of
$m_V/m_{\Lambda_b}$ \cite{SMlambdab}.)

Assuming factorization and using the expressions for $a$, $b$ and $c$
given in Eq.~(\ref{abc}), from Eq.~(\ref{smallmass}) we obtain
\bea
A_0 & = & A_{0X}+A_{0Y} ~, \nn\\
A_{0X} & \approx & 2 m_B m_1 g_{V_1} X \left[ \left( A_1^{(2)} -
A_2^{(2)} \right) + \frac{m_2}{m_B} \left( A_1^{(2)} + A_2^{(2)}
\right) \right] \frac{E^2}{m_1m_2} ~, \nn\\
A_{0Y} & \approx & 2 m_B m_2 g_{V_2} Y \left[ \left( A_1^{(1)} -
A_2^{(1)} \right) + \frac{m_1}{m_B} \left( A_1^{(1)}+A_2^{1} \right)
\right] \frac{E^2}{m_1m_2} ~, \nn\\
A_\| & \approx & -\sqrt{2} m_B \left[ m_1 g_{V_1} \left( 1 + {m_2\over
m_B} \right) A_1^{(2)} (m_1^2) X + m_2 g_{V_2} \left( 1 + {m_1\over
m_B} \right) A_1^{(1)}(m_2^2) Y \right] ~, \nn\\
A_\perp & \approx & -\sqrt{2} m_B \left[ m_1 g_{V_1} \left( 1 -
{m_2\over m_B} \right) V^{(2)}(m_1^2) X \right. \nn\\
&& \hskip2.0truein \left.
+~m_2 g_{V_2} \left( 1 -
{m_2\over m_B} \right) V^{(1)}(m_2^2) Y \right]~.
\eea
The above equations exhibit the same suppression of the $A_{\|,\perp}$
amplitudes relative to $A_0$ as that given in Eq.~(\ref{suppression}).
However, from the expression for $A_0$, one sees that if we have
$A_1\approx A_2$ then the suppression of $A_{\|,\perp}$ relative to
$A_0$ will be diluted from $m_1m_2 / {E^2}$ to simply $m / E$ (where
$m = m_1$ or $m_2$). In fact, this may well be the case: if the
dominant contribution to the form factors comes from soft gluon
interactions between the quarks inside the mesons then one has the
following relations between the vector form factors \cite{Charles}:
\beq
A_1=A_2 +O(m/E) ~~,~~~~ V=A_1+O(m/E) ~.
\label{leet}
\eeq
On the other hand, in the presence of hard gluon interactions, the
relations in the above equation no longer hold. Still, even for this
scenario, the form factors $A_1$ and $A_2$ have been found numerically
to be very similar \cite{Sanda}.

The main point here is that all triple-product correlations in
charmless $B \to V_1 V_2$ decays are suppressed by some power of
$m/E$. We call this {\it mass suppression}.

Thus, all TP's in $B\to V_1 V_2$ decays suffer from a combination of
flavour and mass suppression. These suppressions are most severe for
the $B$ decays to ground-state vector mesons which have been studied
previously. For example, even for interfering amplitudes of similar
size, the flavour suppression from isospin symmetry will produce a
negligible TP. In fact, the earlier calculations \cite{KP} find that
most SM TP's in $B$ decays with final-state ground state mesons are
small, less than $5\%$, and we largely agree with these results (we
have checked these estimates using updated values for the CKM
parameters $\rho$ and $\eta$, the Wilson coefficients, and various
form factors).

There is one point on which we disagree with previous analyses. Some
of the papers in Ref.~\cite{KP} find large [$O(10\%)$] TP's for $B$
decays involving ground-state vector mesons. In general, these large
TP's correspond to decays with final-state $\omega$ mesons. In
principle, such decays might avoid flavour suppression since the
$\omega$ is an $SU(3)$ singlet. However, since the mass, form factors
and decay constants for the $\omega$ are very similar to those of the
$\rho$, flavour suppression is expected to be present even for decays
involving $\omega$'s, and this is what we find through explicit
calculation. We therefore conclude that the TP's for $B$ decays to
ground-state vector mesons are {\it all} small in the SM.

\subsection{SM Triple Products: Radially-Excited Vector Mesons 
and other New Decays}

Based on the discussion above, it is clear that the measurement of TP
asymmetries will be facilitated if one uses the heaviest final-state
vector mesons possible. This will minimize the mass suppression of the
TP's. For example, one could consider decays of the $B_c$ mesons using
$b \to c$ transitions. For charmless $B$ decays it might be more
useful to consider decays to radially-excited states of the vector
mesons $\rho$, $K^*$ or $\phi$. As shown in Ref.~\cite{dattalip1},
such transitions can have branching ratios which may be larger than,
or of the same size as, decays to the ground state configurations.
This is easily understood in the context of factorization.  Consider
the decay $B \rightarrow V_1^{\prime} V_2$ where $V_1^{\prime}$ is the
radially-excited meson and $V_{1,2}$ are the ground-state mesons. We
assume that both $V_1^{\prime}$ and $V_2$ are light mesons. The
amplitude for the process is then
\beq
\bra{V_2(\vec p) \, V_1^{\prime}(-\vec p)}T\ket{B} =
\bra{V_1^{\prime}(-\vec p)}J_{1\mu} \ket{B}\, \bra {V_2(\vec
p)}J^{\mu}_{2} \ket {0} ~,
\eeq
where $J_{1,2}$ are currents that occur in the effective Hamiltonian
[Eq.~(\ref{Heff})]. The transition matrix element for the hadronic
decay can then be written in terms of $B \to V_1^{\prime}$ form
factors and the $V_2$ meson decay constant. The form factors can be
expressed as overlap integrals of the $B$ and the $V_1^{\prime}$ meson
wavefunctions.  When $V_1^{\prime}$ is a light meson, with a mass much
smaller than that of the $B$ meson, the main contributions to the
overlap integrals come from the high-momentum components, or the tail,
of the meson wavefunctions. For a radially-excited meson
$V_1^{\prime}$, which has more high-momentum components (i.e.\ a
longer tail), the overlap integrals will be enhanced compared to those
of the ground-state meson $V_1$.  As a consequence, the $B \to
V_1^{\prime}$ form factors are likely to be increased compared to
those of $B \to V_1$. This would then translate into a larger
branching ratio for $B \to V_1^{\prime} V_2 $ than $B \to V_1 V_2$. In
Ref.~\cite{dattalip1}, this effect was demonstrated explicity with
various confining potentials for the mesons.

Another advantage of using radially-excited mesons is that the TP
asymmetries will not be suppressed by flavour symmetries. For
instance, although the TP in $B^- \to \rho^- \rho^0$ is tiny due to
isospin symmetry, the TP asymmetry in $B^- \to \rho^- \rho^{0\prime}$,
where $\rho^{0\prime}$ is a radially-excited state, does not suffer a
corresponding flavour suppression.

In this section we provide estimates of the sizes of TP's involving
radially-excited vector mesons, as well as several other new modes not
considered before. 

Like any CP-violating signal, TP's will be largest when the two
interfering amplitudes are of comparable size. Also, TP asymmetries
will be maximized when the largest decay amplitude is involved. (If
not, then the denominator will be larger than the numerator, thereby
decreasing the asymmetry.)

With these general ideas in hand, we use the framework of
factorization to estimate numerically the size of the TP's within the
SM. In particular, we use factorization to estimate the matrix
elements of the various operators that appear in the effective
Hamiltonian [Eq.~(\ref{Heff})]. The values of the various Wilson
coefficients are given by \cite{BuraseffH}
\bea
c_{1f} = -0.185 &,& c_{2f} = 1.082 ~, \nn\\
c^t_3 = 0.014 ~,~~ c^t_4 = -0.035 &,& c^t_5 = 0.010 ~,~~
c^t_6 =-0.041 ~, \nn\\
c^t_7 = -1.24\times 10^{-5} ~,~~ c_8^t = 3.77\times 10^{-4} &,&
c_9^t = -0.010 ~,~~ c_{10}^t = 2.06\times 10^{-3} ~.
\eea
Note that the tree operators in Eq.~(\ref{Heff}) can generate via
rescattering the $u$- and $c$-quark penguin pieces, proportional to
$V_{ub}V_{uq}^*$ and $V_{cb}V_{cq}^*$ ($q=d,s$) respectively. The
coefficients associated with the short-distance rescattering effects
are given by
\bea
c_{3,5}^i = -c_{4,6}^i/N_c = P^i_s/N_c &,&
c_{7,9}^i = P^i_e ~,~~
c_{8,10}^i = 0 ~,~~
i=u,c ~,
\label{coeffs}
\eea
where $N_c$ is the number of colours. The leading contributions to
$P^i_{s,e}$ are given by $P^i_s = ({\frac{\alpha_s}{8\pi}}) c_2
({\frac{10}{9}} +G(m_i,\mu,q^2))$ and $P^i_e =
({\frac{\alpha_{em}}{9\pi}}) (N_c c_1+ c_2) ({\frac{10}{9}} +
G(m_i,\mu,q^2))$, in which the function $G(m,\mu,q^2)$ takes the form
\begin{eqnarray}
G(m,\mu,q^2) = 4\int^1_0 x(1-x) \mbox{ln}{m^2-x(1-x)q^2\over
\mu^2} ~\mbox{d}x ~,
\label{rescatt}
\end{eqnarray}
where $q$ is the momentum carried by the virtual gluon in the penguin
diagram. In our calculations, we use a value of $q^2 = m_b^2/2$.

In Sec.~2.3, we presented a list of decays which can yield
triple-product asymmetries in the SM. Below we provide estimates of
the expected size of these TP's for decays which have not been
previously examined. Because we expect that $A_T^{(2)} < A_T^{(1)}$,
the estimates are given for $A_T^{(1)}$ only. We also give the
expected branching ratios for both a given decay and its CP-conjugate
decay. {}From these numbers one can easily obtain the direct CP
asymmetry expected in the decay.

The amplitudes for the various decays depend on combinations of Wilson
coefficients, $a_i$, where $a_i= c_i+{c_{i+1}/ N_c}$ for $i$ odd and
$a_i= c_i+{c_{i-1}/ N_c}$ for $i$ even. The terms described by the
various $a_i$'s can be associated with the different decay topologies
introduced earlier. The terms proportional to $a_2$ and $a_1$ are,
respectively, the colour-allowed and colour-suppressed tree amplitudes
$T$ and $C$. The term proportional to $a_4$ is the colour-allowed
penguin amplitude, $P$, while the terms $a_3$ and $a_5$ represent the
OZI-suppressed amplitudes. Finally, the dominant electroweak penguin
$P_{EW}$ is represented by term proportional to $a_9$, while $a_7$ and
$a_{10}$ are additional small electroweak-penguin amplitudes.

In addition, as mentioned earlier, certain amplitudes can potentially
receive large nonfactorizable corrections. In the past it has been
customary to take into account such nonfactorizable corrections by
treating $N_c$ as a free parameter. In our calculations we adopt the
same prescription, and provide estimates of TP's with two standard
choices: $N_c=3$ (pure factorization) and $N_c= \infty$ (large
nonfactorizable effects included). Now, it is known that $N_c \to
\infty$ is inconsistent with data on $B \to PP$ and $B \to PV$ decays
\cite{chengnonleptonic}. In fact, for charmless $B$ decays the
effective $N_c$ may be different for operators with different chiral
structure in the effective Hamiltonian \cite{chengnonleptonic}. In
this paper we are dealing with $VV$ final states, and the effective
value of $N_c$ which is applicable here will only be known when there
are enough experimental data to carry out a detailed analysis. Our
choice of $N_c=\infty$ can be considered as an extreme case of
nonfactorizable effects. Although it will probably turn out to be
inconsistent with data on nonleptonic $B \to VV$ decays, our purpose
here is simply to be most conservative (perhaps excessively so) in our
estimation of nonfactorizable effects. Realistically, we expect the
true value of the TP to lie somewhere between its values for $N_c=3$
and $N_c= \infty$. In most cases, this allows us to clearly establish
that the TP in question is expected to be small in the SM, even in the
presence of unrealistically large nonfactorizable effects. (A more
complete discussion of nonfactorizable effects can be found in the
next subsection.)

Note that since $c_1$ and $c_2$ have opposite signs, the
colour-suppressed tree amplitude, described by $a_1 = c_1+c_2/N_c$, is
further suppressed because of an accidental cancellation between its
Wilson coefficients. The effect of this suppression depends strongly
on the value taken for $N_c$. For $N_c=3$, one obtains $a_1=0.176$,
while for $N_c= \infty$ we have $a_1=-0.185$, so that even the sign of
the colour-suppressed amplitude is different in the two cases. For the
OZI-suppressed terms $a_{3,5}$ the difference can also be quite
dramatic: for $N_c=3$ we have $a_3=0.002$ and $a_5=-0.0036$, while for
$N_c= \infty$ we have $a_3=0.014$ and $a_5=0.01$. In this latter case
the OZI terms can be of the same order as the colour-allowed penguins.
This fact will be important in understanding the numbers for the
T-violating asymmetries given below. Whether or not the OZI terms are
important in $B$ decays is a matter of debate, and several tests to
find evidence for their presence in $B$ decays have been discussed
recently \cite{dattalip2}.

\subsubsection{$B_c^- \to J/\psi D^{*-}$}

The amplitude for this process is given by
\beq
A[B_c^- \to J/\psi D^{*-}] = \frac{G_F}{\sqrt{2}} [X P_{D^*}+Y
P_{J/\psi}],
\eeq
with
\bea
X & = & V_{cb} V_{cd}^* a_2 - \sum_{q=u,c,t} V_{qb} V_{qd}^* (a_4^q +
a_{10}^q) ~, \nn\\
Y & = & V_{cb} V_{cd}^* a_1 - \sum_{q=u,c,t} V_{qb} V_{qd}^* (a_3^q +
a_5^q + a_7^q + a_9^q) ~, \nn\\
P_{D^*} & = & m_{D^*} g_{D^*} \varepsilon^{*\mu}_{D^*} \bra{J/\psi}
\bar{c} \gamma_{\mu} (1 - \gamma_5) b \ket{B_c^-} ~, \nn\\
P_{J/\psi} & = & m_{J/\psi} g_{J/\psi} \varepsilon^{*\mu}_{J/\psi} 
\bra{D^{*-}} \bar{d} \gamma_{\mu} ( 1-\gamma_5 ) b \ket{B_c^-} ~.
\eea
Note that the main difference between the two amplitudes $X$ and $Y$
is simply the fact that some Wilson coefficients are multiplied by
$1/N_c$ in one amplitude, while they are multiplied by 1 in the
other. This is the case for most of the decays we consider.

We can now calculate $a$, $b$ and $c$ from Eq.~(\ref{abc}) with the
identification $V_1=D^*$ and $V_2=J/\psi$. For numerical results we
will use the following inputs: the CKM parameters are $\rho = 0.17$
and $\eta = 0.39$; the decay constants are $g_{J/\psi}=0.405$ GeV
\cite{PDG} and $g_{D^*}=0.237$ GeV \cite{alaa1}; the form factors for
$B_c^- \to J/\psi$ transitions are given by
$A_1^{(J/\psi)}(m_{D^*}^2)=0.73$, $A_2^{(J/\psi)}(m_{D^*}^2)=0.75$ and
$V^{(J/\psi)}(m_{D^*}^2)=1.1$, while for $B_c^- \to D^{*-}$ they are
$A_1^{(D^*)}(m_{J/\psi}^2)=0.70$, $A_2^{(D^*)}(m_{J/\psi}^2)=1.2$ and
$V^{(D^*)}(m_{J/\psi}^2)=2.1$ \cite{korner}.

\begin{table}[thb] 
\begin{center} 
\begin{tabular}{|c|c|c|c|} 
\hline 
Process & BR & $A_{T}^{(1)}$ (\%) & $ N_{c}$ \\
\hline 
$B_c^-\to J/\psi D^{*-}$ & $3.48~(3.45) \times 10^{-3}$
&0.011 ($-$0.03)& 3 \\
\hline 
$B_c^-\to J/\psi D^{*-}$ & $3.02~(3.0) \times 10^{-3}$ &0.11
($-$0.06)& $\infty$ \\
\hline
\end{tabular}
\end{center}
\caption{Branching ratios (BR) and triple-product asymmetries
($A_{T}^{(1)}$) for $B_c^- \to J/\psi D^{*-}$, for $N_c=3$ (pure
factorization) and $N_c= \infty$ (large nonfactorizable effects). The
results for the CP-conjugate process are given in parentheses. }
\label{Bc}
\end{table}

We present our results in Table~\ref{Bc}, including the branching
ratio and the T-odd triple product $A_T^{(1)}$ for both process and
CP-conjugate process. Regardless of the value taken for $N_c$, the
T-violating asymmetries are expected to be tiny. This is
understandable because, while T-violation comes from $C$--$P$ or
$T$--$P_{EW} (P_{OZI})$ interference, there is a large colour-allowed
tree contribution to the amplitude. Thus, the denominator of
$A_T^{(1)}$ [Eq.~(\ref{TPmeasure})] is always much larger than the
numerator, resulting in a small TP asymmetry. As expected, the
T-violation is larger for $N_c = \infty$ because of enhanced OZI
terms, but the asymmetries are still too small to be measurable.

\subsubsection{$B^- \to \rho^{0\prime} K^{*-}$, $\rho^{0} K^{*-\prime}$,
$\omega^{\prime} K^{*-}$, $\omega K^{*-\prime}$}

The amplitude for $B$ decays to radially-excited vector mesons has a
similar form to that for $B$ decays to ground-state vector mesons.  We
therefore start with the decay amplitude for ground-state mesons.  For
$V=\rho^0$ or $\omega$, this amplitude can be written as
\beq
A[B^- \to K^{*-} V] = \frac{G_F}{\sqrt{2}} [X_{V}
P_{K^*}^{V}+Y_V P_{V}^{K^{*}}],
\label{BKminusVamp1}
\eeq
with
\bea
X_{\rho} = X_{\omega} & = & V_{ub}V_{us}^* a_2 - \sum_{q=u,c,t} V_{qb}
V_{qs}^* \left( a_4^q + a_{10}^q \right) ~, \nn\\
Y_{\rho} & = & V_{ub} V_{us}^* a_1 - \sum_{q=u,c,t} V_{qb} V_{qs}^*
\left( \frac{3}{2} a_7^q + \frac{3}{2} a_9^q \right) ~, \nn\\
Y_{\omega} & = & V_{ub} V_{us}^* a_1 - \sum_{q=u,c,t} V_{qb} V_{qs}^*
\left( 2 a_3^q + 2 a_5^q + \frac{1}{2} a_7^q + \frac{1}{2} a_9^q
\right) ~, \nn\\
P_{K^*}^{\rho} & = & m_{K^*}g_{K^*}\varepsilon^{*\mu}_{K^*}
\bra{\rho^0} \bar{u} \gamma_{\mu}(1-\gamma_5)b \ket{B^-} ~, \nn\\
P_{\rho^0}^{K^{*}} & = & m_{\rho^0}g_{\rho^0}
\varepsilon^{*\mu}_{\rho^0} \bra{K^{*-}} \bar{s}
\gamma_{\mu}(1-\gamma_5)b \ket{B^-} ~, \nn\\
P_{K^*}^{\omega} & = & m_{K^*}g_{K^*}\varepsilon^{*\mu}_{K^*}
\bra{\omega} \bar{u} \gamma_{\mu}(1-\gamma_5)b \ket{B^-} ~, \nn\\
P_{\omega}^{K^{*}} & = & m_{\omega} g_{\omega}
\varepsilon^{*\mu}_{\omega} \bra{K^{*-}} \bar{s}
\gamma_{\mu}(1-\gamma_5)b \ket{B^-} ~.
\label{BKminusVamp}
\eea
The decays which interest us involve $\rho(1450)$, $\omega(1420)$
and/or $K^{*}(1410)$ in the final state. Quark-model predictions
classify these states as radially-excited states of the $\rho$,
$\omega$ and the $K^{*}$. Henceforth we will label these states as
$\rho^{0\prime}$, $\omega'$ and $K^{* \prime}$. The amplitude for a
decay involving an excited final state can be obtained simply from
Eqs.~(\ref{BKminusVamp1}) and (\ref{BKminusVamp}) by replacing
$\rho^0$, $\omega$ and/or $K^*$ by $\rho^{0\prime}$, $\omega'$ and/or
$K^{* \prime}$.

To calculate the TP asymmetries for these decays, we need the form
factors for transitions of a $B$-meson to such radially-excited
states. These are obtained by assuming a linear confining potential
for the light mesons, and the wavefunction $N e^{-p^2/p_F^2}$ with the
fermi momentum $p_F=0.3$ ($0.5$) GeV for $\bd$ ($\bs$) mesons
\cite{dattalip1}. The results for the form factors are
\bea
{A_1^{\rho^{\prime}}(q^2=m_{\rho}^2)\over
A_1^{\rho}(q^2=m_{\rho^{\prime}}^2)} &=& 1.38 ~, \nn\\
{A_2^{\rho^{\prime}}(q^2=m_{\rho}^2)\over
A_2^{\rho}(q^2=m_{\rho^{\prime}}^2)} & = & 1.2 ~, \nn\\
{V^{\rho^{\prime}}(q^2=m_{\rho}^2)\over
V^{\rho}(q^2=m_{\rho^{\prime}}^2)} &=& 1.66 ~.
\label{radial-rho}
\eea 
We assume the same values for the ratios of form factors for $B \to
\omega'$ and $B \to K^{*\prime}$ transitions. This is reasonable since
any $SU(3)$ breaking effects should cancel in the ratios of form
factors. Note that Eq.~(\ref{radial-rho}) is given in terms of the form
factors for ground-state mesons. Identifying $V_1=K^*$ and $V_2=
\rho^0$ or $\omega$, for $B^- \to \rho^0$ these are
$A_1^{(\rho)}(m_{K^*}^2)=0.26$, $A_2^{(\rho)}(m_{K^*}^2)=0.24$ and
$V_1^{(\rho)}(m_{K^*}^2)=0.31$, while for $B^- \to K^{*-}$ transitions
the form factors are given by $A_1^{(K^*)}(m_\rho^2)=0.36$,
$A_2^{(K^*)}(m_{\rho}^2)=0.32$ and $V^{(K^*)}(m_{\rho}^2)=0.44$
\cite{mstech}.  We assume that the form factors for $B \to \omega$ are
the same as for $B \to \rho^0$.

An estimate of the triple products for these decays also involves the
decay constants of the radially-excited vector mesons. These are found
to be similar to the decay constants for the ground state mesons. We
take $f_{\rho} = f_{\rho^{\prime}} = f_{\omega} = f_{\omega'} = 0.190$
GeV and $f_{K^*} = f_{K^{*\prime}} = 0.214$ GeV.

There are two advantages to using a final state with one
radially-excited vector meson: the $m_V/m_B$ suppression is reduced,
and there is no flavour symmetry relating the final-state vector
mesons which would result in a further suppression. One could also
consider final states containing two radially-excited states. In this
case, however, the suppression due to flavour symmetry would apply
again and for this reason we do not examine these decays here. One
could also consider final states with radially-excited and
orbitally-excited vector mesons. This would require the calculation of
form factors for transitions of $B$ mesons to orbitally-excited
states. This interesting possibility is beyond the scope of this work
and will be investigated elsewhere. As we will see, for final states
with a single radially-excited state measurable TP asymmetries may be
possible, and this should encourage a more thorough study of TP's in
$B$ decays to radially and orbitally-excited vector mesons.

Based on Eq.~(\ref{BKminusVamp}) we can make the following
observations.  For the decays $B^- \to \rho^{0\prime} K^{*-}$ and $B^-
\to \rho^0 K^{*-\prime}$, the interfering amplitudes are of unequal
size, which further suppresses the TP asymmetry. For $B^- \to
\omega^{\prime} K^{*-}$ and $B^- \to \omega K^{*-\prime}$, there is a
possibility of an enhanced OZI contribution for $N_c = \infty$.  This
could interfere with the colour-allowed tree which is suppressed by
CKM factors.
 
The results of Table~\ref{BKminusV} are consistent with these
observations. All TP's are expected to be very small with two
exceptions: for $N_c= \infty$ a measurable TP asymmetry is predicted
for $B^-\to \omega^{\prime} K^{*-} $ (8\%) and possibly $B^-\to \omega
K^{*-\prime} $ (4\%). These TP asymmetries are generated mainly from
$T$--$P_{OZI}$ interference (the $C$--$P$ interference is
small). Thus, it is possible that nonfactorizable effects can generate
large TP asymmetries for these decays to radially-excited states. We
should stress, however, that this is far from guaranteed -- the $N_c =
\infty$ prescription is just an estimate. The true nonfactorizable
effects could be much smaller than this. Conversely, for other decays,
it appears unlikely that such effects can lead to measurable TP's --
these decays are therefore excellent places to search for new physics.

\begin{table}[thb] 
\begin{center} 
\begin{tabular}{|c|c|c|c|} 
\hline 
Process &  BR & $A_T^{(1)}$ \% & $ N_{c}$ \\ 
\hline 
$B^-\to \rho^{0\prime} K^{*-}$ & $11.8~(7.9) \times 10^{-6}$ & 0.53
 ($-$0.13) & 3 \\
\hline 
$B^-\to \rho^{0\prime} K^{*-}$ & $13.1~(10.0) \times 10^{-6}$ & 1.2
 (0.45) & $\infty$ \\
\hline
$B^-\to \rho^{0} K^{*-\prime}$ & $10~(7.0) \times 10^{-6}$ &$-$0.87
 (0.21) & 3 \\
\hline 
$B^-\to \rho^{0} K^{*-\prime}$ & $11.1~(9.1) \times 10^{-6}$ &$-$1.9
 ($-$0.68) & $\infty$ \\
\hline
$B^-\to \omega^{\prime} K^{*-}$ & $10~(6) \times 10^{-6}$ &0.16
 ($-$0.29) & 3 \\
\hline 
$B^-\to \omega^{\prime} K^{*-}$ & $2.3~(3.6) \times 10^{-6}$ &$-$10.6
($-$5.2) & $\infty$ \\
\hline
$B^-\to \omega K^{*-\prime}$ & $7.7~(4.7) \times 10^{-6}$ &$-$0.29
 (0.51) & 3 \\
\hline 
$B^-\to \omega K^{*-\prime}$ & $6.5~(8.8) \times 10^{-6}$ & 5.4
(3.0) & $\infty$ \\
\hline 
\end{tabular}
\end{center}
\caption{Branching ratios (BR) and triple-product asymmetries
($A_{T}^{(1)}$) for $B^- \to \rho^{0\prime} K^{*-}$, $\rho^{0}
K^{*-\prime}$, $\omega^{\prime} K^{*-}$ and $\omega K^{*-\prime}$, for
$N_c=3$ (pure factorization) and $N_c= \infty$ (large nonfactorizable
effects). The results for the CP-conjugate process are given in
parentheses.}
\label{BKminusV}
\end{table}

\subsubsection{$\bdbar\to K^{*0} \rho^{0\prime}$, 
$K^{*0 \prime} \rho^{0}$, $K^{*0} \omega^{\prime}$, $K^{*0\prime}
\omega$}

As before, we first present the amplitude for the ground-state mesons.
For the decays $\bdbar \to K^{*0} V$ with $V= \rho^0$ or $\omega$, the
amplitude is given by
\beq
A[\bdbar \to K^{*0} V] =
\frac{G_F}{\sqrt{2}} [X_{V} P_{K^*}^{V}+Y_V P_{V}^{K^{*}}]~,
\eeq
with
\bea
X_{\rho} = - X_\omega & = & \sum_{q=u,c,t} V_{qb} V_{qs}^*
\left( a_4^q + a_{10}^q \right) ~, \nn\\
Y_{\rho} & = & V_{ub}V_{us}^* a_1 - \sum_{q=u,c,t} V_{qb} V_{qs}^*
\left( \frac{3}{2} a_7^q + \frac{3}{2} a_9^q \right) ~, \nn\\
Y_{\omega} & = & V_{ub} V_{us}^* a_1 - \sum_{q=u,c,t} V_{qb} V_{qs}^*
\left( 2 a_3^q + 2 a_5^q + \frac{1}{2} a_7^q + \frac{1}{2} a_9^q
\right) ~, \nn\\
P_{K^*}^{\rho} & = & m_{K^*}g_{K^*}\varepsilon^{*\mu}_{K^*}
\bra{\rho^0} \bar{d} \gamma_{\mu}(1-\gamma_5)b \ket{\bdbar} ~, \nn\\
P_{\rho^0}^{K^{*}} & = & m_{\rho^0} g_{\rho^0}
\varepsilon^{*\mu}_{\rho^0} \bra{K^{*0}} \bar{s}
\gamma_{\mu}(1-\gamma_5)b \ket{\bdbar} , \nn\\
P_{K^*}^{\omega} & = & m_{K^*}g_{K^*}\varepsilon^{*\mu}_{K^*}
\bra{\omega} \bar{d} \gamma_{\mu}(1-\gamma_5)b \ket{\bdbar} ~, \nn\\
P_{\omega}^{K^{*}} & = & m_{\omega} g_{\omega}
\varepsilon^{*\mu}_{\omega} \bra{K^{*0}} \bar{s}
\gamma_{\mu}(1-\gamma_5)b \ket{\bdbar} ~.
\eea
Again, to obtain the amplitude for a decay involving an excited final
state, one simply replaces $\rho^0$, $\omega$ and/or $K^*$ by
$\rho^{0\prime}$, $\omega'$ and/or $K^{* \prime}$ in the above
equation.

Above, we have used the flavour wavefunction $\rho^0 = (\bar{u} u -
\bar{d} d)/\sqrt{2}$ and $\omega = (\bar{u} u + \bar{d} d)/ \sqrt{2}$,
and similarly for the excited states. The effect of the relative sign
in the flavour wavefunctions of the $\rho^0$ and $\omega$ has been
included in the definition of the phases $X_{\rho, \omega}$ and
$Y_{\rho, \omega}$, and so does not have to be included in the $B \to
\rho^0 (\omega)$ form factors. We shall follow this convention in
subsequent decays involving $\rho^0$ and $\omega$.

The triple-product correlations for these decays are presented in
Table~\ref{BK0V}. The TP's are from $C$--$P$ interference, which is
small, so that we do not find large TP's in this case. It is only in
the decay $\bdbar\to K^{*0} \omega'$ that a marginally
measurable TP ($\sim 3$--4\%) might be found. However this again
relies on large nonfactorizable effects, which may or may not be
present.

\begin{table}[thb] 
\begin{center} 
\begin{tabular}{|c|c|c|c|} 
\hline 
Process &  BR & $A_T^{(1)}$ \% & $ N_{c}$ \\ 
\hline 
$\bdbar\to K^{*0} \rho^{0\prime}$ & $11.3~(10.9) \times 10^{-6}$ &
 0.05 ($-$0.61)& 3 \\
\hline 
$\bdbar\to K^{*0} \rho^{0\prime}$ & $13.5~(14.6) \times 10^{-6}$ &
 0.64 ($-$0.14)& $\infty$ \\
\hline
$\bdbar\to K^{*0 \prime} \rho^{0}$ & $9.7~(9.3) \times 10^{-6}$
 &$-$0.08 (1.0)& 3 \\
\hline 
$\bdbar\to K^{*0 \prime} \rho^{0}$ & $12.0~(12.9) \times 10^{-6}$
 &$-$1.0 (0.23)& $\infty$ \\
\hline
$\bdbar\to K^{*0} \omega^{\prime}$ & $9.1~(8.6) \times 10^{-6}$
 &$-$0.14 ($-$0.52)& 3 \\
\hline 
$\bdbar\to K^{*0} \omega^{\prime}$ & $1.0~(0.7) \times 10^{-6}$ & 3.8
 (3.13)& $\infty$ \\
\hline
$\bdbar\to K^{*0\prime} \omega$ & $7.0~(10) \times 10^{-6}$
 & 0.26 (0.65)& 3 \\
\hline 
$\bdbar\to K^{*0\prime} \omega$ & $5.5~(4.9) \times 10^{-6}$
 &$-$1.0 ($-$0.73)& $\infty$ \\
\hline
\end{tabular}
\end{center}
\caption{Branching ratios (BR) and triple-product asymmetries
($A_{T}^{(1)}$) for $\bdbar\to K^{*0} \rho^{0\prime}$, $K^{*0 \prime}
\rho^{0}$, $K^{*0} \omega^{\prime}$ and $K^{*0\prime} \omega$, for
$N_c=3$ (pure factorization) and $N_c= \infty$ (large nonfactorizable
effects). The results for the CP-conjugate process are given in
parentheses.}
\label{BK0V}
\end{table}

\subsubsection{$B^-\to \rho^{-} \rho^{0\prime}$,
$\rho^{-\prime} \rho^{0}$, $\rho^{-} \omega'$, $\rho^{-\prime}
\omega$}

The amplitude for the ground-state decay $B^- \to \rho^- V$ with $
V=\rho^0$ or $\omega$ is given by
\beq
A[B^- \to \rho^- V ] = \frac{G_F}{\sqrt{2}} [X_{V}
P_{\rho^-}^{V}+Y_V P_{V}^{\rho^{-}}]~,
\eeq
with
\bea
X_{\rho} = X_{\omega} & = & V_{ub}V_{ud}^* a_2 - \sum_{q=u,c,t} V_{qb}
V_{qd}^* \left( a_4^q+a_{10}^q \right) ~, \nn\\
Y_{\rho} & = & V_{ub} V_{ud}^* a_1 - \sum_{q=u,c,t} V_{qb} V_{qd}^*
\left( - a_4^q + \frac{3}{2} a_7^q + \frac{3}{2} a_9^q + \frac{1}{2}
a_{10}^q \right) ~, \nn\\
Y_{\omega} & = & V_{ub} V_{ud}^* a_1 - \sum_{q=u,c,t} V_{qb} V_{qd}^*
\left( a_4^q + 2 a_3^q + 2 a_5^q + \frac{1}{2} a_7^q + \frac{1}{2}
a_9^q - \frac{1}{2} a_{10}^q \right) ~, \nn\\
P_{\rho^-}^{\rho^0} & = & m_{\rho^-} g_{\rho^-}
\varepsilon^{*\mu}_{\rho^-} \bra{\rho^0} \bar{u}
\gamma_{\mu}(1-\gamma_5)b \ket{B^-} ~, \nn\\
P_{\rho^0}^{\rho^{-}} & = &\frac{1}{\sqrt{2}} m_{\rho^0} g_{\rho^0}
\varepsilon^{*\mu}_{\rho^0} \bra{\rho^-} \bar{d} \gamma_{\mu}
(1-\gamma_5)b \ket{B^-} ~, \nn\\
P_{\rho^-}^{\omega} & = & m_{\rho^-} g_{\rho^-}
\varepsilon^{*\mu}_{\rho^-} \bra{\omega} \bar{u}
\gamma_{\mu}(1-\gamma_5)b \ket{B^-} ~, \nn\\
P_{\omega}^{\rho^{-}} & = &\frac{1}{\sqrt{2}} m_{\omega} g_{\omega}
\varepsilon^{*\mu}_{\omega} \bra{\rho^-} \bar{d} \gamma_{\mu}
(1-\gamma_5) b \ket{B^-} ~.
\eea
The amplitude for a decay involving an excited final state is obtained
by replacing $\rho^0$ and/or $\omega$ by $\rho^{0\prime}$ and/or
$\omega'$.

In this case, the TP's arise mostly from $T$--$P$ interference,
neither of which is CKM-suppressed. However, the penguin amplitude $P$
is only about 4\% of the tree amplitude $T$, so that the maximum TP
asymmetry turns out to be small, $\sim 1$ \%. The results of our
calculations are presented in Table~\ref{BrhominusV}. Note that, if
one assumes isospin conservation, flavour suppression leads to an
identically vanishing TP for the ground-state decay $B^-\to \rho^-
\rho^0$.

\begin{table}[thb] 
\begin{center} 
\begin{tabular}{|c|c|c|c|} 
\hline 
Process &  BR & $A_T^{(1)}$ \% & $ N_{c}$ \\ 
\hline 
$B^-\to \rho^{-} \rho^{0\prime}$ & $29.3~(29.7) \times 10^{-6}$ &0.51
 (0.33)& 3 \\
\hline 
$B^-\to \rho^{-} \rho^{0\prime}$ & $18.3~(18.7) \times 10^{-6}$ &0.66
 (0.41)& $\infty$ \\
\hline
$B^-\to \rho^{-\prime} \rho^{0}$ & $23.8~(23.5) \times 10^{-6}$
 &$-$0.63 ($-$0.42)& 3 \\
\hline 
$B^-\to \rho^{-\prime} \rho^{0}$ & $12~(11.7) \times 10^{-6}$ &$-$1.0
 ($-$0.65)& $\infty$ \\
\hline
$B^-\to \rho^{-} \omega'$ & $27.4~(32.7) \times 10^{-6}$
 &$-$0.58 ($-$0.38)& 3 \\
\hline 
$B^-\to \rho^{-} \omega'$  & $17.9~(19.8) \times 10^{-6}$
 &0.1 (0.04)& $\infty$ \\ 
\hline
$B^-\to \rho^{-\prime} \omega$ & $15.8~(20.0) \times 10^{-6}$ &$-$1.16
 ($-$0.72)& 3 \\
\hline 
$B^-\to \rho^{-\prime} \omega$ & $4.0~(4.2) \times 10^{-6}$ &0.49
 (0.22)& $\infty$ \\
\hline  
\end{tabular}
\end{center}
\caption{Branching ratios (BR) and triple-product asymmetries
($A_{T}^{(1)}$) for $B^-\to \rho^{-} \rho^{0\prime}$, $\rho^{-\prime}
\rho^{0}$, $\rho^{-} \omega'$ and $\rho^{-\prime} \omega$, for $N_c=3$
(pure factorization) and $N_c= \infty$ (large nonfactorizable
effects). The results for the CP-conjugate process are given in
parentheses.}
\label{BrhominusV}
\end{table}

\subsubsection{$\bdbar\to \rho^{0} \omega'$,
$\rho^{0\prime} \omega$}

The amplitude for the ground-state decay $\bdbar \to \rho^0 \omega$ is
given by
\beq
A[\bdbar \to \rho^0 \omega] = \frac{G_F}{\sqrt{2}} [X P_{\rho^0}+Y
P_{\omega}] ~,
\eeq
with
\bea
X & = & V_{ub} V_{ud}^* a_1 + \sum_{q=u,c,t} V_{qb} V_{qd}^* \left(
a_4^q + \frac{3}{2} a_7^q + \frac{3}{2} a_9^q \right) ~,\nn\\
Y & = & -V_{ub} V_{ud}^* a_1 + \sum_{q=u,c,t} V_{qb} V_{qd}^* \left(
a_4^q + 2 a_3^q + 2 a_5^q + \frac{1}{2} a_7^q + \frac{1}{2} a_9^q +
\frac{1}{2} a_{10}^q \right) ~, \nn\\
P_{\rho^0} & = & \frac{1}{\sqrt{2}} m_{\rho^0} g_{\rho^0}
\varepsilon^{*\mu}_{\rho^0} \bra{\omega} \bar{d}
\gamma_{\mu}(1-\gamma_5)b \ket{\bdbar} ~, \nn\\
P_{\omega} & = &\frac{1}{\sqrt{2}} m_{\omega} g_{\omega}
\varepsilon^{*\mu}_{\omega} \bra{\rho^0} \bar{d}
\gamma_{\mu}(1-\gamma_5)b \ket{\bdbar} ~.
\eea

The TP's in this case are due principally to $C$--$P$ interference.
Neither of these amplitudes is CKM-suppressed, and they are of similar
size. As a consequence, while the TP's for the ground-state decay are
small, due to flavour and mass suppressions, we find measurable
asymmetries for decays with radially-excited vector mesons in the
final state (see Table~\ref{Brho-omega}). Unfortunately, the branching
ratios for all these decays are expected to be in the $10^{-7}$
range. Furthermore, the TP asymmetry changes sign as $N_c$ is varied
from 3 to $\infty$.  Thus, there is again no guarantee of a large TP
-- it is possible that nonfactorizable effects are such that the
actual TP is small.

\begin{table}[thb] 
\begin{center} 
\begin{tabular}{|c|c|c|c|} 
\hline 
Process &  BR & $A_T^{(1)}$ \% & $ N_{c}$ \\ 
\hline 
$\bdbar\to \rho^{0} \omega'$ & $4.5~(1.8) \times 10^{-7}$ &6.2
 (10.2)& 3 \\
\hline 
$\bdbar\to \rho^{0} \omega'$ & $0.5~(0.52) \times 10^{-7}$
 &$-$17.2 ($-$11.1)& $\infty$ \\
\hline
$\bdbar\to \rho^{0\prime} \omega$ & $6~(3.3) \times 10^{-7}$ &6.0
 (6.3)& 3 \\
\hline 
$\bdbar\to \rho^{0\prime} \omega$ & $2.45~(2.08) \times 10^{-7}$
 &$-$4.0 ($-$3.2)& $\infty$ \\
\hline
\end{tabular}
\end{center}
\caption{Branching ratios (BR) and triple-product asymmetries
($A_{T}^{(1)}$) for $\bdbar\to \rho^{0} \omega'$ and $\rho^{0\prime}
\omega$, for $N_c=3$ (pure factorization) and $N_c= \infty$ (large
nonfactorizable effects). The results for the CP-conjugate process are
given in parentheses. }
\label{Brho-omega}
\end{table}

\subsubsection{$\bsbar\to \phi^{\prime} K^{*}$,
$\phi K^{* \prime}$}

We are also interested in the pure $b \to d$ penguin decay $\bsbar \to
\phi^{(\prime)} K^{*(\prime)}$, where $\phi^{\prime}$ corresponds to
the radially-excited state $\phi(1680)$. For the form factors for
these decays, we obtain
\bea
{A_1^{\phi^{\prime}}(q^2=m_{K^*}^2)\over A_1^{\phi}(q^2=m_{K^{*}}^2)}
&=& 1.5 ~, \nn\\
{A_2^{\phi^{\prime}}(q^2=m_{K^*}^2)\over A_2^{\phi}(q^2=m_{K^{*}}^2)}
&=& 1.35 ~, \nn\\
{V^{\phi^{\prime}}(q^2=m_{K^*}^2)\over V^{\phi}(q^2=m_{K^{*}}^2)}
&=& 1.8 ~. \nn\\
\label{radial-phi}
\eea
The form factors for the ground-state transitions can be found in
Ref.~\cite{mstech}. For the decay constants we use $f_{\phi^{\prime}}
= f_{\phi} = 0.237$ GeV.

The amplitude for the ground-state decay $\bsbar \to \phi K^{*0}$ is
given by
\beq
A[\bsbar \to \phi K^{0*}] = \frac{G_F}{\sqrt{2}} [X P_{K^*}+Y
P_{\phi}]~,
\eeq
with
\bea
X & = & - \sum_{q=u,c,t} V_{qb} V_{qd}^* \left( a_4^q -\frac{1}{2}
a_{10}^q \right) ~, \nn\\
Y & = &- \sum_{q=u,c,t} V_{qb} V_{qd}^* \left( a_3^q + a_5^q -
\frac{1}{2} a_7^q - \frac{1}{2} a_9^q \right) ~, \nn\\
P_{K^*} & = & m_{K^*}g_{K^*}\varepsilon^{*\mu}_{K^*} \bra{\phi}
\bar{s} \gamma_{\mu}(1-\gamma_5)b \ket{\bsbar} ~, \nn\\
P_{\phi} & = & m_{\phi}g_{\phi}\varepsilon^{*\mu}_{\phi} \bra{K^{0*}}
 \bar{d} \gamma_{\mu}(1-\gamma_5)b \ket{\bsbar} ~.
\eea

In this case, the TP's arise mainly from $P$--$P_{EW}(P_{OZI}$
interference. We find a marginally measurable TP asymmetry only for
$\bsbar \to \phi K^{*\prime}$ with $N_c= \infty$, i.e.\ with enhanced
OZI terms, and the branching ratio for this decay is tiny,
$O(10^{-8})$. Our results are presented in Table~\ref{Bsphi-K}.

\begin{table}[thb] 
\begin{center} 
\begin{tabular}{|c|c|c|c|} 
\hline 
Process &  BR & $A_T^{(1)}$ \% & $ N_{c}$ \\ 
\hline 
$\bsbar\to \phi^{\prime} K^{*}$ & $11~(5.5) \times 10^{-7}$ &$-$0.17
 (0.21)& 3 \\
\hline 
$\bsbar \to \phi^{\prime} K^{*}$ & $2.8~(1.3) \times 10^{-7}$ &$-$1.14
 (1.51)& $\infty$ \\
\hline
$\bsbar\to \phi K^{* \prime}$ & $6.3~(3.1) \times 10^{-7}$ &0.23
 ($-$0.31)& 3 \\
\hline 
$\bsbar \to \phi K^{*\prime}$ & $0.15~(0.06) \times 10^{-7}$ &16.8
($-$22.9)& $\infty$ \\
\hline 
\end{tabular}
\end{center}

\caption{Branching ratios (BR) and triple-product asymmetries
($A_{T}^{(1)}$) for $\bsbar\to \phi^{\prime} K^{*}$ and $\phi K^{*
\prime}$, for $N_c=3$ (pure factorization) and $N_c= \infty$ (large
nonfactorizable effects). The results for the CP-conjugate process are
given in parentheses. }

\label{Bsphi-K}
\end{table}

\subsubsection{$B^- \to D^{*0} K^{*-}$, ${\bar D}^{*0} K^{*-}$; 
$B^- \to D^{*0} \rho^-$, ${\bar D}^{*0} \rho^-$}

We now examine $B^-$ decays in which the final-state $D^{*0}$ or
${\bar D}^{*0}$ mesons subsequently decay to the same state. We assume
that $D^{0*} \to D^0 \pi^0$ and ${\bar D}^{0*} \to {\bar D}^0 \pi^0$,
with $D^0, {\bar D}^0 \to f$, where $f = K^+ \pi^-$ or $f = \pi^+
\pi^-$.

Consider first $B^- \to D^{*0} K^{*-}$ and $B^- \to {\bar D}^{*0}
K^{*-}$. The decay amplitude is given by
\beq
A[B^- \to K^{*-} f] = \frac{G_F}{\sqrt{2}} [X P_{K^*}+Y P_{D^*}] ~,
\eeq
with
\bea
X & = & V_{cb} V_{us}^* a_2 \sqrt{B_1}\sqrt{B_2} ~, \nn\\
Y & = & V_{cb}V_{us}^*a_1\sqrt{B_1}\sqrt{B_2} + V_{ub} V_{cs}^* a_1
 \sqrt{B_1}\sqrt{B_2^{\prime}} ~, \nn\\
P_{K^*} & = & m_{K^*}g_{K^*}\varepsilon^{*\mu}_{K^*} \bra{{\bar
D}^{0*}} \bar{c} \gamma_{\mu}(1-\gamma_5)b \ket{B^-} ~, \nn\\
P_{D^*} & = & m_{D^*}g_{D^*}\varepsilon^{*\mu}_{D^*} \bra{K^{*-}}
 \bar{s} \gamma_{\mu}(1-\gamma_5)b \ket{B^-} ~,
\eea 
where $B_1$ is the branching ratio for $D^{0*} \to D^0 \pi^0$ and
$B_2(B_2^{\prime})$ are the branching ratios for $D^0 ({\bar D}^0) \to
f$. The values for the form factors for $B \to D^*$ transitions are
$A_1(m_{K^*}^2) = V(m_{K^*}^2) = 0.783$ and $A_2(m_{K^*}^2) = 0.772$
\cite{alaa2}. (Due to heavy quark symmetry, the form factors have very
similar values.) Now, the relative strong phase between the amplitudes
$D^0 \to f$ and ${\bar D}^0 \to f$ is unknown. In our estimates, we
choose this phase to be zero. This assumption is not unreasonable
since these transitions go through colour-allowed tree decays, so that
any strong phases generated by nonfactorizable effects are likely to
be small.

\begin{table}[thb] 
\begin{center} 
\begin{tabular}{|c|c|c|c|} 
\hline 
Process &  BR & $A_T^{(1)}$ \% & $ N_{c}$ \\ 
\hline 
$B^-\to K^{*-} (f= K^+ \pi^-)$ & $3.5~(3.5) \times 10^{-7}$ &4.1
 (4.1)& 3 \\
\hline 
$B^-\to K^{*-} (f= K^+ \pi^-)$ & $1.4~(1.4) \times 10^{-7}$ &$-$11.0
 ($-$11.0)& $\infty$ \\
\hline
$B^-\to K^{*-} (f= \pi^+ \pi^-)$ & $10.2~(10.2) \times 10^{-7}$ &0.52
 (0.52)& 3 \\
\hline 
$B^-\to K^{*-} (f= \pi^+ \pi^-)$ & $5.8~(5.8) \times 10^{-7}$ &$-$1.03
 ($-$1.03)& $\infty$ \\
\hline 
$B^-\to \rho^- (f= K^+ \pi^-)$ & $18.0~(18.0) \times 10^{-7}$ &$-$0.56
 ($-$0.56)& 3 \\
\hline 
$B^-\to \rho^- (f= K^+ \pi^-)$ & $14.0~(14.0) \times 10^{-7}$ &0.81
 (0.81)& $\infty$ \\
\hline
$B^-\to \rho^- (f= \pi^+ \pi^-)$ & $18.2~(18.2) \times 10^{-6}$
 &$-$0.03 ($-$0.03)& 3 \\
\hline 
$B^-\to \rho^- (f= \pi^+ \pi^-)$ & $13.1~(13.1) \times 10^{-6}$ &0.04
 (0.04)& $\infty$ \\
\hline
\end{tabular}
\end{center}
\caption{Branching ratios (BR) and triple-product asymmetries
($A_{T}^{(1)}$) for $B^- \to D^{*0} K^{*-}$ and $B^- \to {\bar D}^{*0}
K^{*-}$, as well as $B^- \to D^{*0} \rho^-$ and $B^- \to {\bar D}^{*0}
\rho^-$, for $N_c=3$ (pure factorization) and $N_c= \infty$ (large
nonfactorizable effects). It is assumed that $D^0, {\bar D}^0 \to f$,
with $f = K^+ \pi^-$ or $f = \pi^+ \pi^-$. The results for the
CP-conjugate process are given in parentheses. }
\label{B-Ddecays}
\end{table}

We present our results in Table~\ref{B-Ddecays}. We find that the
T-violating asymmetries may be measurable for the decays with $f = K^+
\pi^-$, but are small for $f = \pi^+ \pi^-$. These results can be
understood as follows. The decay $B^- \to D^{*0} K^{*-}$ is dominated
by a colour-allowed tree diagram ($T$) and involves the CKM matrix
elements $V_{cb} V_{cd}^*$, while $B^- \to {\bar D}^{*0} K^{*-}$ is
colour-suppressed ($C$) and involves $V_{ub} V_{cs}^*$. Thus, these two
amplitudes are of very different size -- the latter is roughly 5\% of
the former. However, in order to obtain a sizeable TP, it is necessary
to have two decay amplitudes of similar magnitudes. This can occur if
the decays $D^0 \to f$ and ${\bar D}^{0} \to f$ are, respectively,
doubly-Cabibbo-suppressed and Cabibbo-allowed, which is the case for
$f=K^+ \pi^-$. (This is similar to the method for obtaining $\gamma$
proposed in Ref.~\cite{ADS}.) Unfortunately, the net branching ratio
is small $O(10^{-7})$. On the other hand, for $f = \pi^+ \pi^-$, both
the $D^0$ and ${\bar D}^{0}$ decays are singly-Cabibbo-suppressed, so
the TP is small.

We now turn to the decays $B^- \to D^{*0} \rho^-$ and $B^- \to {\bar
D}^{*0} \rho^-$. The amplitude in this case given by
\beq
A[B^- \to \rho^{-} f] = \frac{G_F}{\sqrt{2}} [X P_{\rho^-}+Y P_{D^*}] ~,
\eeq
with
\bea
X & = & V_{cb} V_{ud}^* a_2 \sqrt{B_1}\sqrt{B_2} ~, \nn\\
Y & = & V_{cb}V_{ud}^*a_1\sqrt{B_1}\sqrt{B_2} + V_{ub} V_{cd}^* a_1
\sqrt{B_1}\sqrt{B_2^{\prime}} ~, \nn\\
P_{\rho^-} & = & m_{\rho^-}g_{\rho^-}\varepsilon^{*\mu}_{\rho^-} \bra{{\bar
D}^{0*}} \bar{c} \gamma_{\mu}(1-\gamma_5)b \ket{B^-} ~, \nn\\
P_{D^*} & = & m_{D^*}g_{D^*}\varepsilon^{*\mu}_{D^*} \bra{\rho^-}
\bar{d} \gamma_{\mu}(1-\gamma_5)b \ket{B^-} ~.
\eea
In this case, the second decay ($B^- \to {\bar D}^{*0} \rho^-$) is
also suppressed relative to the first ($B^- \to D^{*0} \rho^-$).
However, here the suppression is much larger than in $B^- \to D^{*0}
K^{*-}$, ${\bar D}^{*0} K^{*-}$ decays -- in addition to the ratio
$C/T$, there is also a suppression due to the ratio of CKM matrix
elements, $\vert V_{ub} V_{cd}^* / V_{cb} V_{ud}^* \vert$. Thus,
regardless of the final state $f$ in $D^0, {\bar D}^0 \to f$, the two
amplitudes remain very different in size, leading to small TP's. This
expectation is borne out in Table~\ref{B-Ddecays}.

\subsubsection{$B_c^- \to {\bar D}^{*0} D_s^{*-}$, $D^{*0}
   D_s^{*-}$; $B_c^- \to {\bar D}^{*0} D^{*-}$, $D^{*0} D^{*-}$}

Finally, we consider pairs of $B_c^-$ decays to final states including
$D^{*0}$ or ${\bar D}^{*0}$ mesons. Unfortunately, there are no
calculations yet of the form factors for $B_c^- \to {\bar D}^{*0}$,
$D_s^{*-}$ and $D^{*-}$ transitions. As a result, we can only present
``back-of-the-envelope'' estimates of the triple products for these
decays. (Still, based on our analyses of the previous decays, these
estimates are probably reasonably accurate.)

Consider first $B_c^- \to {\bar D}^{*0} D_s^{*-}$, $D^{*0} D_s^{*-}$.
The decay $B_c^- \to {\bar D}^{*0} D_s^{*-}$ is dominated by $T$ and
involves $V_{ub} V_{cs}^*$, while $B_c^- \to D^{*0} D_s^{*-}$ is
governed by $C$ and $V_{cb} V_{us}^*$. The two amplitudes are
therefore comparable in size, which naively suggests that one can
obtain a measurable TP by using decays such as $D^0, {\bar D}^0 \to
\pi^+\pi^-$, which are both singly-Cabibbo-suppressed. However, note
that, within factorization, the two $B_c^-$ decay amplitudes are
proportional to $f_{B_c^- \to {\bar D}^{*0}} f_{D_s^{*-}}$ and
$f_{B_c^- \to D_s^-} f_{D^{*0}}$, which are related by flavour $SU(3)$
symmetry. We therefore expect the TP asymmetries to be small for these
decays. However, the TP's could be measurable if one uses final states
involving excited mesons.

The situation is similar for $B_c^- \to {\bar D}^{*0} D^{*-}$, $D^{*0}
D^{*-}$. In this case, the amplitude for the second $B_c^-$ decay is
actually larger than the first (by about a factor of 10). Thus, in
order to obtain roughly equal overall amplitudes, one has to use
doubly-Cabibbo-suppressed decays such as $D^0, {\bar D}^0 \to K^+
\pi^-$. However, even in this case one expects tiny TP asymmetries:
the two $B_c^-$ decay amplitudes are proportional to $f_{B_c^- \to
{\bar D}^{*0}} f_{D^{*-}}$ and $f_{B_c^- \to D^{*-}} f_{D^{*0}}$,
which are related by isospin. The only way to obtain measurable TP's
is if the final states involve excited mesons.

\subsection{Nonfactorizable effects}

In our analysis, we have used factorization to calculate the expected
size of triple-product asymmetries in certain $B \to V_1 V_2$
decays. We have included potential nonfactorizable contributions by
considering also the case $N_c= \infty$ in the $a_i$ (which are
combinations of the Wilson coefficients and $N_c$). In this
subsection, we examine in more detail nonfactorizable effects. In
particular, we are interested in establishing which decays are likely
to be most (and least) affected by such effects. We also explore the
properties of those nonfactorizable effects which can modify the TP
predictions. The determination of which TP predictions are the most
reliable in turn indicates which decay modes are best to use in the
search for new physics.

The most interesting decays are those for which the TP asymmetries are
predicted to be very small (or zero) in the framework of factorization
within the SM. If it can be established that nonfactorizable effects
do not significantly affect these predictions, the measurement of a
sizeable nonzero TP asymmetry would clearly signal the presence of new
physics. In such decays, within factorization, we can expresses the
various linear polarization amplitudes in the following form:
\bea
A_i & = & R_i [ P_1 + P_2 e^{i\phi} e^{i\Delta} ] = R_i X ~,
\label{withoutnonfac}
\eea  
where $i=0$, $\|$ and $\perp$. The weak and strong phases are denoted
by $\phi$ and $\Delta$, respectively, while the $R_i$ are real numbers
that depend on form factors and decay constants. The quantities
$P_{1,2}$ depend on combinations of the Wilson coefficients and the
magnitude of the CKM elements, and are therefore real. With the above
parametrization, it is clear that there are no TP asymmetries, since
all amplitudes have the same phase, i.e.\ ${\rm Im}[A_0 A_\perp^*] =
{\rm Im}[A_\| A_\perp^*] \sim {\rm Im}[XX^*]=0$.

TP asymmetries can potentially be generated in such decays if
nonfactorizable effects are present. One possibility is that there are
additional contributions, such as annihilation diagrams, which
contribute to the decay. In this case, the new amplitude can interfere
with the amplitude in Eq.~(\ref{withoutnonfac}) to generate a TP
asymmetry. The full decay amplitude then has the general form
\beq
A_i = R_i X + R_i^{\prime} Y ~,
\label{withann}
\eeq    
where $X$ and $Y$ depend differently on the weak and the strong
phases. In practice, however, such annihilation contributions are
suppressed in the heavy-quark limit. The annihilation terms can be
estimated in the framework of QCD factorization \cite{BBNS}. For $VV$
final states, the annihilation terms are not chirally enhanced, unlike
$PP$ and $PV$ states \cite{ChengYang}. Thus, these contributions are
purely power suppressed ($ \sim O(1/m_b) $) in the heavy-quark
expansion, and are small.

Another class of nonfactorizable effects are those which modify the
individual $P_{1,2}$ amplitudes in Eq.~(\ref{withoutnonfac}). The
general form of the amplitudes is then
\beq
A_i = R_i[P_1(1+a_ie^{i\alpha_i})+P_2(1+b_ie^{i\beta_i})e^{i\phi}
e^{i\Delta}] ~,
\label{withnonfac}
\eeq  
where $\alpha_i$ and $\beta_i$ are strong phases generated by the
nonfactorizable effects. Note that if the quantities $a_i$, $b_i$,
$\alpha_i$ and $\beta_i$ are the same for all three linear
polarization amplitudes, then the TP asymmetries will still vanish,
even in the presence of nonfactorizable effects. Thus, it is only
nonfactorizable contributions that affect the amplitudes $A_0$, $A_\|$
and $A_\perp$ differently which can generate a TP asymmetry. 

One can see explicitly how a TP is generated by nonfactorizable
effects by rewriting the $A_i$ in Eq.~(\ref{withnonfac}):
\beq
A_i = R_i X_i + R_i Y_i ~,
\eeq
with
\bea
X_i & = & (P_1+P_2e^{i\phi}e^{i\Delta})(1+a_ie^{i \alpha_i}) \nn\\
    &= & X(1+a_ie^{i \alpha_i}) ~, \nn\\
Y_i & = & P_2(b_ie^{i\beta_i}-a_ie^{i\alpha_i})e^{i \phi}e^{i\Delta} ~.
\eea
The interference of $X_i$ with $Y_i$ (specifically, $P_1$--$Y_i$
interference) leads to a TP. (Note that the interference of two
different $X_i$'s does not lead to a true T-violating effect since
$X$, the term containing the weak-phase information, is the same for
all three amplitudes.) Thus, not only must the nonfactorizable effects
be different for the three $A_i$, but the nonfactorizable corrections
to $P_1$ and $P_2$ should also be different. If this were not the
case, then the $Y_i$ would vanish. 

We have therefore seen that TP's can be generated by the interference
of factorizable and nonfactorizable contributions. This then begs two
questions: (i) which amplitudes are most likely to be affected by
nonfactorizable effects, and (ii) how big are such effects? The first
question is easy to answer: contributions which are suppressed in the
factorization framework, such as the colour-suppressed tree amplitude
$C$, are likely to receive large nonfactorizable contributions. This
was already seen in the previous subsection in which we parametrized
nonfactorizable effects by varying $N_c$. The value and the sign of
$a_1 = c_1+c_2/N_c$, which describes $C$, depend strongly on the value
chosen for $N_c$. Thus, TP asymmetries that arise from the
interference of colour-allowed and colour-suppressed transitions, as
happens for several of the decay modes, can be significantly modified
by nonfactorizable effects.

The second question is more difficult to answer. Various methods to
calculate nonfactorizable effects have been considered recently in the
literature, but there is no compelling evidence for the validity of
any one approach. For example, in QCD factorization \cite{BBNS},
nonfactorizable effects can be different for the different helicity
states (linear polarization states) \cite{ChengYang}. However, some of
these corrections, such as the hard spectator corrections, are
dominated by soft configurations and turn out to be divergent. Hence
nothing quantitative can be said about the size of TP asymmetries
generated by these nonfactorizable corrections. Still, it should be
noted that qualitatively these corrections are quite significant for
colour-suppressed amplitudes while the fractional change in
colour-allowed amplitudes is proportional to $\alpha_s(m_b) \sim 0.2$
and is small. It is very likely that the measurement of triple-product
asymmetries in $B\to V_1 V_2$ decays will provide useful information
about the dynamics of nonleptonic decays.

The conclusion here is that the most reliable predictions for TP's are
for those decays where both $P_1$ and $P_2$ are colour allowed. There
are many examples of these. For example, the decays $\bdbar \to D^{*+}
D^{*-}$, $B^- \to D^{*0} D^{*-}$, $\bsbar \to D_s^{*+} D^{*-}$,
$\bdbar \to \rho^+ \rho^-$, $\bsbar \to K^{*+} \rho^-$, $B_c^- \to
{\bar D}^{*0} \rho^-$ have both colour-allowed tree and colour-allowed
$b\to d$ penguin contributions; $\bdbar \to K^{*-} \rho^+$, $\bsbar
\to K^{*+} K^{*-}$, $B_c^- \to {\bar D}^{*0} K^{*-}$ have both
colour-allowed tree and colour-allowed $b \to s$ penguin contributions;
$\bdbar \to K^{*0} {\bar K}^{*0}$, $B^- \to K^{*0} K^{*-}$, $B_c^- \to
K^{*0} D^{*-}$ are pure colour-allowed $b \to d $ penguin decays.
Within factorization, the TP asymmetries in all of these decays are
expected to vanish, even though there are two decay amplitudes
($P_{1,2}$) with a relative weak phase. Since the nonfactorizable
effects in these decays are expected to be small, any measurement of
large T-violating triple products in these decays will be a clear
signal of new physics. (Of course, one can add to this list processes
which are dominated by a single amplitude, such as $B \to J/\psi K^*$
($b \to c \bar{c} s$) or $B \to \phi K^*$ (pure $b \to s $ penguin),
since no TP asymmetries are expected in these decays.)

\subsection{Discrete Ambiguities}

Although most of the triple-product asymmetries are predicted to be
small in the decays we have studied, a handful of TP's may be
measurable. What can we learn from them? The answer is that, apart
from testing our knowledge of hadronic $B$ decays, these TP's can
potentially be used to remove an important discrete ambiguity in the
measurements of the CP angles of the unitarity triangle.

Within the SM, CP violation is signalled by nonzero values of
$\alpha$, $\beta$ and $\gamma$, the three internal angles of the
so-called unitarity triangle (UT) \cite{PDG}. By measuring
CP-violating rate asymmetries in the $B$ system, one can obtain the
three CP angles $\alpha$, $\beta$ and $\gamma$. Any inconsistency
among the angles and sides of the UT indicates the presence of new
physics. The standard decays used for obtaining these CP angles are
$\bd(t) \to J/\psi \ks$ ($\beta$), $\bd(t) \to \pi^+ \pi^-$
($\alpha$), and $B^\pm \to D K^\pm$ ($\gamma$) \cite{CPreview}.
Unfortunately, these decays only allow the extraction of $\sin
2\alpha$, $\sin 2\beta$ and $\sin^2 \gamma$, which leads to a fourfold
ambiguity for each of the angles. 

If one assumes that the three angles add up to $180^\circ$, which
holds even in the presence of new physics in $\bd$--$\bdbar$ mixing
\cite{NirSilv}, then most of the values for the angle sets
$(\alpha,\beta,\gamma)$ are forbidden. However, one still has a
twofold ambiguity in the construction of the UT \cite{KayLon}. Given
that $\sin 2\beta$ has been measured to be positive, there are two
scenarios: (i) if $\sin(2 \alpha) > 0$, then both UT's point up, while
(ii) if $\sin(2 \alpha) < 0$, then one UT points up, while the other
points down. In either scenario, if one of the solutions is consistent
with the SM, while the other is not, then it is necessary to resolve
this discrete ambiguity in order to be certain that new physics is
present.

Consider now a decay for which the TP is predicted to be large. As
noted in the introduction, the TP is proportional to $\sin \phi \cos
\delta$, where $\phi$ and $\delta$ are weak and strong phases,
respectively. In fact, it is straightforward to show that all TP's are
proportional to the CKM parameter $\eta$. $\eta$ measures the height
of the UT, so that if $\eta > 0$ ($< 0$), the UT points up (down).
Therefore, the {\it sign} of the TP can tell us whether the UT points
up or down, thus resolving the discrete ambiguity in the second
scenario above.

Unfortunately, things are not quite so easy. Once again, one needs to
understand well the nonfactorizable effects. For example, consider the
decay $\bdbar\to \rho^{0} \omega'$ (Table~\ref{Brho-omega}).
If $N_c = 3$ (pure factorization), then the TP asymmetry is predicted
to be $+8\%$, while if $N_c= \infty$ (large nonfactorizable effects
included), the asymmetry is $-14\%$. Suppose, then, that an asymmetry
of $-8\%$ is measured. This could imply one of two things: either (i)
there are no nonfactorizable effects, but $\eta <0$, or (ii)
nonfactorizable effects are important, and $\eta > 0$. Unless one can
distinguish theoretically between these two possibilities, the
measurement of the TP asymmetry will not tell us whether the UT points
up or down. Thus, TP's can potentially resolve the above discrete
ambiguity, but significant theoretical input will probably be
required.

\section{New Physics}

In almost all of the decays we have studied, the triple-product
asymmetries are predicted to be very small, so that these are good
places to search for physics beyond the SM. In this section, we
examine in more detail the kinds of new physics which can generate
such TP's.

Consider $B\to V_1 V_2$ decays which have only one kinematical
amplitude in the standard model (or for which one such amplitude
dominates). Because there is only a single amplitude, no T-violating
TP asymmetry can be produced. Now, as we saw earlier, the effective SM
Hamiltonian involves only a left-handed $b$-quark, and so contains
only $(V-A) \times (V-A)$ and $(V-A) \times (V+A)$ operators. However,
some types of new physics can couple to the right-handed $b$-quark,
producing $(V+A) \times (V-A)$ and/or $(V+A) \times (V+A)$
operators. These new-physics operators will produce different
kinematical amplitudes, leading to different phases for $a$, $b$ and
$c$, and giving rise to a TP asymmetry.

This can be seen explicitly as follows. Suppose that only $B\to V_2$
transitions occur in the decay $B\to V_1 V_2$. The SM contribution to
such a decay, $A_{SM}$, is given in Eq.~(\ref{firstterm}), repeated
here for convenience:
\beq
A_{SM} \sim \sum_{{\cal O},{\cal O}'} \bra{V_1} {\cal O} \ket{0}
\bra{V_2} {\cal O}' \ket{B} = X \varepsilon_1^{*\mu} \bra{V_2} {\bar
q}' \gamma_\mu (1 - \gamma_5) b \ket{B} ~.
\label{firsttermSM}
\eeq
Recall that all weak-phase information is contained in the factor $X$.
Now assume that there is a new-physics contribution with a $(V+A)
\times (V-A)$ or $(V+A) \times (V+A)$ structure. The new amplitude is
then
\beq
A_{NP} \sim \sum_{{\cal O_{NP}},{\cal O_{NP}}'} \bra{V_1} {\cal
O_{NP}} \ket{0} \bra{V_2} {\cal O_{NP}}' \ket{B} = Y
\varepsilon_1^{*\mu} \bra{V_2} {\bar q}' \gamma_\mu (1 + \gamma_5) b
\ket{B} ~,
\label{firsttermNP}
\eeq
where $Y$ contains the new-physics weak phase information. In the
presence of the new-physics contribution the amplitudes $a$, $b$ and
$c$ of Eq.~(\ref{abcdefs}) can now be written as
\bea
a &=& -m_B m_1 g_{V_1} A_1^{(2)}(m_1^2)
\left(1 + {m_2 \over m_B} \right ) 
\left[ 
X -  Y \right] ~, \nn\\
b &=& 2 m_B m_1 g_{V_1} A_2^{(2)}(m_1^2) 
\left(1 + {m_2 \over m_B} \right )^{-1} 
\left[  X -  Y \right] ~,
\nn\\
c &=& - m_B m_1 g_{V_1} V^{(2)}(m_1^2)
 \left(1 + {m_2 \over m_B} \right )^{-1} 
\left[  X +  Y \right] ~.
\label{abcNP}
\eea 
Thus, when the new-physics contributions are included, ${\rm Im}(a
c^*)$ and ${\rm Im}(b c^*)$ are nonzero. That is, a TP asymmetry will
arise due to the interference of $X$ and $Y$. Furthermore, since the
SM and new-physics operators have different structures, there is no
flavour symmetry relating the two contributions, i.e.\ the phases of
$a$, $b$ and $c$ are different even if $V_1 = V_2$. That is, in the
presence of new physics there is no suppression of the TP asymmetry
due to flavour symmetries. Note also that these TP asymmetries can be
generated by the interference of two colour-allowed amplitudes (most
TP's in the SM are due to the interference of a colour-allowed and a
colour-suppressed amplitude).

We therefore see that the measurement of a nonzero TP asymmetry in
this class of decays would be a smoking-gun signal for the presence of
nonstandard operators, specifically those involving a right-handed
$b$-quark \cite{RHNP}. In fact, as was shown in Ref.~\cite{NPlambdab},
by studying TP's in several such modes, one can test various models of
new physics. As an example of this, we concentrate on the decay $B \to
\phi K^*$.

Within the SM, the CP asymmetries in both $\bd(t) \to J/\psi \ks$ and
$\bd(t) \to \phi\ks$ are expected to measure $\sin 2\beta$. Any
differences between these two measurements should be at most at the
level of O($\lambda^2$), where $\lambda \sim 0.2$. However, at present
there appears to be an inconsistency. The world averages for these
measurements are \cite{Jpsiks,phiks}:
\bea
\sin (2 \beta (J/\psi \ks)) & = & 0.734 \pm 0.054 ~, \nn\\
\sin (2 \beta (\phi \ks)) & = & -0.39 \pm 0.41 ~.
\label{phiKsresult}
\eea
Now, decays that have significant penguins contributions are most
likely to be affected by physics beyond the SM \cite{Gross}. In
particular, it was pointed out some time ago that $\bd \to \phi \ks$
is sensitive to new physics because it is a pure $b\to s$ penguin
decay \cite{LonSoni}. For this reason, there have been several recent
papers discussing possible new-physics scenarios which can account for
the above discrepancy \cite{phiKsNP,dattarparity,bhaskar}. (Some of
these have sought a simultaneous explanation of the CP asymmetry
measurements in $\bd(t) \to \phi \ks$ and the $B \to \eta^{\prime} K$
branching ratios \cite{bhaskar}. However, it should be pointed out
that the SM explanation of these branching ratios is far from being
ruled out \cite{dattalip3}.)

Assuming that there is physics beyond the SM in $\bd \to \phi \ks$,
the question then is: what is the nature of this new physics? More
concretely, what is the structure of the new-physics operators that
contribute to the effective Hamiltonian for $B$ decays? A partial
answer to this question can be found in the measurement of T-violating
triple products in the sister decay $B \to \phi K^*$ \cite{LSS2}. As
we have argued above, if TP's vanish in certain decays in the SM, they
can be generated in models of new physics which involve couplings to
the right-handed $b$-quark. However, not all of the models proposed to
explain the CP asymmetry in $\bd(t) \to \phi \ks$ contain such
couplings. One can therefore partially distinguish among these models
by examining TP's in $B \to \phi K^*$. (Note that one can also look at
TP's in $\Lambda_b \to \Lambda \phi$ \cite{NPlambdab} as the
underlying $ b \to s {\bar{s}}s$ transition in this decay is the same
as in $B \to \phi \ks$.)

We do not present here a comprehensive analysis, but rather focus on
one particular new-physics model, that of supersymmetry with R-parity
violation \cite{dattarparity}. (Note that the analysis here can easily
be extended to a more general approach, in which one examines
new-physics operators without reference to a particular model. Such an
approach was presented in Ref.~\cite{NPlambdab}.) Assuming that
R-parity-violating SUSY is the explanation for the CP measurements in
$\bd(t) \to \phi \ks$, we estimate here the expected TP asymmetries in
$B \to \phi K^*$.

For the $b \to s {\bar{s}}s$ transition, the relevant terms in the
R-parity-violating SUSY Lagrangian are
\beq
L_{eff} = \frac{\lambda^{\prime}_{i32} \lambda^{\prime*}_{i22}} { 4
m_{ \widetilde{\nu}_i}^2} \bar s (1+\gamma_5) s \, \bar {s}
(1-\gamma_5) b+ \frac{\lambda^{\prime}_{i22} \lambda^{\prime*}_{i23}}
{ 4 m_{ \widetilde{\nu}_i}^2} \bar s (1-\gamma_5) s \, \bar {s}
(1+\gamma_5) b ~.
\eeq
(We refer to Ref.~\cite{dattarparity} for a full explanation of SUSY
with R-parity violation.) The amplitude for $B \to \phi K^{*}$,
including the new-physics contributions, can then be written as
\beq
A[B \to \phi K^*] = \frac{G_F}{\sqrt{2}} [(X+X_1) P_{\phi}+X_2
Q_{\phi}] ~,
\eeq
with
\bea
X & = & - \sum_{q=u,c,t}V_{qb}V_{qs}^* \left[ a_3^q+a_4^q+a_5^q
  -\frac{1}{2}(a_7^q+a_9^q+a_{10}^q) \right] ~, \nn\\
X_1 &= & -\frac{\sqrt{2}}{G_F} \frac{\lambda^{\prime}_{i32}
\lambda^{\prime*}_{i22}} {24 \, m_{ \widetilde{\nu}_i}^2} ~, \nn\\
X_2 & = & -\frac{\sqrt{2}}{G_F} \frac{\lambda^{\prime}_{i22}
  \lambda^{\prime*}_{i23}} {24 \, m_{ \widetilde{\nu}_i}^2} ~, \nn\\
P_{\phi} & = & m_{\phi}g_{\phi}\varepsilon^{*\mu}_{\phi} \bra{K^*}
 \bar{s} \gamma_{\mu}(1-\gamma_5)b \ket{B} ~, \nn\\
Q_{\phi} & = & m_{\phi}g_{\phi}\varepsilon^{*\mu}_{\phi} \bra{K^*}
 \bar{s} \gamma_{\mu}(1+\gamma_5)b \ket{B} ~.
\label{Bphi-K}
\eea
For $\bd \to \phi \ks$ it is the combination $X_1+X_2$ which
contributes \cite{dattarparity}, and we can define the quantity $X_R$
via
\beq
X_1+X_2=\frac{\sqrt{2}}{G_F}\frac{X_R}{12M^2}e^{i\phi} ~,
\eeq
where $\phi$ is the weak phase in the R-parity-violating couplings,
and $M$ is a mass scale with $M \sim m_{\widetilde{\nu}_i}$. In order
to reproduce the CP-violating $\bd(t) \to \phi \ks$ measurement in
Eq.~(\ref{phiKsresult}), one requires $|X_R| \sim 1.5 \times 10^{-3}$
for $M=100$ GeV, along with a value for the phase $\phi$ near ${\pi
\over 2}$. In our calculations of TP's in $B \to \phi K^{*}$ we make
the simplifying assumption that $X_1=X_2$, and choose $\phi={\pi \over
2}$.

We present our results in Table.~\ref{Bdphi-K}. Note that these hold
for both neutral and charged $B$ decays. The branching ratio for $B
\to \phi K^*$ is slightly larger than the measured branching ratios
$BR(B^+ \to \phi K^{*+}) = 10^{+5}_{-4} \times 10^{-6}$ and $BR(\bd
\to \phi K^{*0}) = 9.5^{2.4}_{-2.0} \times 10^{-6}$ \cite{PDG}, but it
is well within the theoretical uncertainties of the calculation. The
important observation is that we expect {\it very large} (15--20\%) TP
asymmetries for these decays, as well as for those with
radially-excited final states.

In fact, these results are not unique to supersymmetry with R-parity
violation. One expects to find large TP asymmetries in many other
models of physics beyond the SM. The measurement of such TP
asymmetries would not only reveal the presence of new physics, but
more specifically it would point to new physics which includes large
couplings to the right-handed $b$-quark.

There is one final point which must be stressed here. The standard
method of searching for new physics in such decays is to try to
measure direct CP asymmetries. However, here such asymmetries are
small, at most 4\%. The reason is simply that direct CP asymmetries
are proportional to $\sin\delta$, where $\delta$ is the strong phase
difference between the two decay amplitudes [Eq.~(\ref{Adirform})],
and for this set of decays the strong phase difference is very
small. Indeed, this is the case for many $B$ decays. This emphasizes
the importance of measuring triple-product asymmetries in order to
search for physics beyond the SM. If one relies only on direct CP
asymmetries, it is easy to miss the new physics.

\begin{table}[thb] 
\begin{center} 
\begin{tabular}{|c|c|c|c|} 
\hline 
Process &  BR & $A_T^{(1)}$ \% & $ N_{c}$ \\ 
\hline 
$B \to \phi K^{*}$ & $16.7~(17.4) \times 10^{-6}$ &$-$16.3 ($-$15.6)&
 3 \\
\hline 
$B \to \phi^{\prime} K^{*}$ & $19.1~(20.7) \times 10^{-6}$ &$-$21.0
 ($-$19.3)& 3 \\
\hline 
$B \to \phi K^{* \prime}$ & $28.0~(28.9) \times 10^{-6}$ &$-$15.4
 ($-$14.8)& 3 \\
\hline 
\end{tabular}
\end{center}
\caption{Branching ratios (BR) and triple-product asymmetries
($A_{T}^{(1)}$) for $B \to \phi K^{*}$ and excited states, for $N_c=3$
(pure factorization). The results for the CP-conjugate process are
given in parentheses.}
\label{Bdphi-K}
\end{table}

\section{Summary}

A great deal of work, both theoretical and experimental, has been
devoted to the study of CP violation in the $B$ system. As always, the
hope is that we will discover physics beyond the standard model. Most
of this work has concentrated on indirect CP-violating asymmetries,
while a smaller fraction has focussed on direct CP violation. However,
one subject which has been largely neglected is T-violating
triple-product correlations (TP's) which take the form $\vec v_1 \cdot
(\vec v_2 \times \vec v_3)$, where each $v_i$ is a spin or
momentum. One point we have attempted to emphasize in this paper is
that TP's are an excellent way to look for new physics.

The idea is straightforward. If one measures a nonzero value for a
quantity which is expected to vanish in the SM, one will clearly have
found new physics. Now, direct CP asymmetries are proportional to
$\sin\phi \sin\delta$, where $\phi$ and $\delta$ are, respectively,
weak and strong phase differences. In $B$ decays, the strong phases
are expected to be small in general, so that the direct CP asymmetries
will be unmeasurable. Note that weak-annihilation contributions
induced by $(S-P)(S+P)$ penguin operators can lead to large strong
phases in certain $PP$ and $PV$ decays \cite{PQCD}, leading to
measurable direct CP asymmetries. However, the annihilation amplitude
in the $VV$ case does {\it not} gain a chiral enhancement of order
$m_B^2/(m_qm_b)$ -- it is truly power suppressed in the heavy-quark
limit \cite{ChengYang}. Hence, in $B \to VV$ decays, the strong phases
are expected to be small, so that direct CP asymmetries will be tiny.

These strong phases will also be small in the presence of new
physics. This is because the new-physics amplitude is typically
expected to be of the same size as loop amplitudes in the SM, and so
any rescattering effects from these new operators will be small,
resulting in small strong phases. (Note that in the SM the strong
phases for $B \to VV$ decays arise dominantly from rescattering of
tree-level amplitudes.) Furthermore, even though the new-physics
contribution may contain different short-distance physics than that of
the SM, the process of hadronization to the final-state mesons is a
QCD phenomena, and so is expected to be same with or without new
physics. Hence, if the SM strong phases are small in $B \to VV$
decays, they are likely to be small even with new physics.  Thus, even
if new physics is present, it will probably be undetectable in $B \to
VV$ decays using direct CP asymmetries.

On the other hand, triple-product asymmetries are proportional to
$\sin\phi \cos\delta$, which are maximized if $\delta \simeq 0$. Thus,
if a TP is predicted to vanish in the SM, this is an excellent place
to look for new physics because there is no suppression from the
strong phases. In particular, if new physics is present, it will be
detected in TP's but not in direct CP asymmetries.

In this paper we have examined in detail triple-product asymmetries in
$B \to V_1 V_2$ decays \cite{Valencia,KP}. It is well known that one
can perform an angular analysis on such decays (usually to separate
the final state into CP-even and CP-odd pieces). However, it is rarely
emphasized that the TP's are in fact the coefficients of some of the
terms in the angular analysis. Thus, the TP asymmetries can be
obtained from such an analysis.

Within factorization, there are relatively few $B \to V_1 V_2$ decays
which are expected to have TP's. The point is that it is not enough to
have two decay amplitudes with a relative weak phase (e.g.\ a tree and
penguin amplitude) -- what one really needs are two {\it kinematical}
amplitudes with a relative phase. In particular, because the SM
interactions are purely left-handed, both $B \to V_1$ and $B\to V_2$
transitions must be allowed. This strongly limits the number of decays
in which TP's are expected, which helps in the search for new physics.

Like previous analyses \cite{Valencia,KP}, we have found several $B$
decays which satisfy these criteria. However, there are two factors
which can suppress the TP's in such decays. First, if $V_1$ and $V_2$
are related by a symmetry such as isospin or $SU(3)$ flavour, the TP
asymmetry is suppressed by the size of symmetry breaking. It is
therefore best to use decays in which the two final-state vector
mesons are unrelated by such a symmetry. Second, all TP's are
suppressed by at least one power of $m_V/m_B$, so that it is best to
use heavy final-state mesons. The upshot is that it is advantageous to
consider decays which involve excited mesons in the final state. In
such decays, the above suppressions are minimized (and the branching
ratios are expected to be of the same size as those involving
ground-state mesons). In this paper, we have therefore concentrated
mainly on decays with radially-excited vector mesons. We have also
considered new modes involving $B_c^-$ decays, as well as $B$ decays
to ${\bar D}^{*0}$ and ${ D}^{*0}$ which then decay to the same final
state.

For those decays which can have nonzero triple products in the SM, we
have calculated the expected size of these TP's. We have found that
most TP's are very small. The only processes where large TP's ($>5\%$)
can occur are in $B$ decays to excited final-state vector mesons,
specifically $B^-\to K^{*-} \omega'$, $\bdbar\to \rho^{0} \omega'$ and
$\bdbar\to \rho^{0\prime} \omega$. Decays with TP's of several percent
(i.e.\ only marginally measurable) include $B^-\to K^{*-\prime}
\omega$, $\bdbar\to K^{*0} \omega'$ and $\bsbar \to \phi K^{*\prime}$.
We have also considered $B$ decays to final states which include
$D^{*0}$ or ${\bar D}^{*0}$ mesons, in which these mesons decay to the
same final state. Large TP's are possible only for $B^-\to K^{*-}
D^{*0}$ and $B^-\to K^{*-}{\bar D}^{*0}$, with $D^{0*} \to D^0 \pi^0$
and ${\bar D}^{0*} \to {\bar D}^0 \pi^0$, and $D^0, {\bar D}^0 \to K^+
\pi^-$.

Note that the sizes of these TP's all depend on the size of the
nonfactorizable effects. In particular, if large TP's are not found in
these decays, it does not necessarily indicate new physics -- it could
simply be that the nonfactorizable effects are such that the TP's are
small.

The most reliable estimates of TP's are for those decays in which
nonfactorizable effects are expected to be small. These are decays
which are dominated by colour-allowed transitions. As it turns out,
most TP's in such decays are expected to vanish, so that these are
excellent processes in which to search for physics beyond the standard
model. As an example of how new physics can affect triple products, we
considered a supersymmetric model with R-parity violation, and
calculated the size of TP's in $B\to \phi K^*$ decays. In the SM, the
TP for this decay vanishes, but when the new-physics contribution is
added, {\it very large} TP's are obtained, in the range 15\%--20\%.
Indeed, this type of result is expected in many models of new
physics. The measurement of a nonzero TP asymmetry where none is
expected would not only reveal the presence of new physics, but more
specifically it would point to new physics which includes large
couplings to the right-handed $b$-quark. This illustrates quite
clearly the usefulness of triple-product correlations in $B$ decays
for finding new physics.

\bigskip
\noindent
{\bf Acknowledgements}:
%\bigskip
D.L. thanks A.D. for the hospitality of the University of Toronto,
where part of this work was done. This work was financially supported
by NSERC of Canada. 

%%%%%%%%%%%%%%%%%%%%% REFERENCES %%%%%%%%%%%%%%%%%%%%%%%%%%%%%%%%


\begin{thebibliography}{99}
\bibitem{RHNP} The importance of new-physics couplings to the
  right-handed $b$ quark was also discussed in D. Atwood and A. Soni,
  Ref.~\cite{KP}.
\end{thebibliography}
\end{document}